\documentclass[twoside,twocolumn,9pt]{article}
\usepackage{extsizes}
\usepackage[super,sort&compress,comma]{natbib} 
\usepackage[version=3]{mhchem}
\usepackage[left=1.5cm, right=1.5cm, top=1.785cm, bottom=2.0cm]{geometry}
\usepackage{balance}
\usepackage{mathptmx}
\usepackage{sectsty}
\usepackage{graphicx} 
\usepackage{lastpage}
\usepackage[format=plain,justification=justified,singlelinecheck=false,font={stretch=1.125,small,sf},labelfont=bf,labelsep=space]{caption}
\usepackage{float}
\usepackage{fancyhdr}
\usepackage{fnpos}
\usepackage[english]{babel}
\addto{\captionsenglish}{%
  
}
\usepackage{array}
\usepackage{droidsans}
\usepackage{charter}
\usepackage[T1]{fontenc}
\usepackage[usenames,dvipsnames]{xcolor}
\usepackage{setspace}
\usepackage[compact]{titlesec}
\usepackage{hyperref}
\usepackage{comment}

\usepackage{epstopdf}

\definecolor{cream}{RGB}{222,217,201}

\begin{document}

\pagestyle{fancy}
\thispagestyle{plain}
\fancypagestyle{plain}{
\renewcommand{\headrulewidth}{0pt}
}

\makeFNbottom
\makeatletter
\renewcommand\LARGE{\@setfontsize\LARGE{15pt}{17}}
\renewcommand\Large{\@setfontsize\Large{12pt}{14}}
\renewcommand\large{\@setfontsize\large{10pt}{12}}
\renewcommand\footnotesize{\@setfontsize\footnotesize{7pt}{10}}
\makeatother

\renewcommand{\thefootnote}{\fnsymbol{footnote}}
\renewcommand\footnoterule{\vspace*{1pt}%
\color{cream}\hrule width 3.5in height 0.4pt \color{black}\vspace*{5pt}} 
\setcounter{secnumdepth}{5}

\makeatletter 
\renewcommand\@biblabel[1]{#1}            
\renewcommand\@makefntext[1]%
{\noindent\makebox[0pt][r]{\@thefnmark\,}#1}
\makeatother 
\renewcommand{\figurename}{\small{Fig.}~}
\sectionfont{\sffamily\Large}
\subsectionfont{\normalsize}
\subsubsectionfont{\bf}
\setstretch{1.125} 
\setlength{\skip\footins}{0.8cm}
\setlength{\footnotesep}{0.25cm}
\setlength{\jot}{10pt}
\titlespacing*{\section}{0pt}{4pt}{4pt}
\titlespacing*{\subsection}{0pt}{15pt}{1pt}

\fancyfoot{}
\fancyfoot[LO,RE]{\vspace{-7.1pt}\includegraphics[height=9pt]{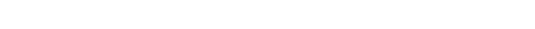}}
\fancyfoot[CO]{\vspace{-7.1pt}\hspace{13.2cm}\includegraphics{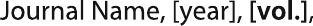}}
\fancyfoot[CE]{\vspace{-7.2pt}\hspace{-14.2cm}\includegraphics{RF.pdf}}
\fancyfoot[RO]{\footnotesize{\sffamily{1--\pageref{LastPage} ~\textbar  \hspace{2pt}\thepage}}}
\fancyfoot[LE]{\footnotesize{\sffamily{\thepage~\textbar\hspace{3.45cm} 1--\pageref{LastPage}}}}
\fancyhead{}
\renewcommand{\headrulewidth}{0pt} 
\renewcommand{\footrulewidth}{0pt}
\setlength{\arrayrulewidth}{1pt}
\setlength{\columnsep}{6.5mm}
\setlength\bibsep{1pt}

\makeatletter 
\newlength{\figrulesep} 
\setlength{\figrulesep}{0.5\textfloatsep} 

\newcommand{\topfigrule}{\vspace*{-1pt}%
\noindent{\color{cream}\rule[-\figrulesep]{\columnwidth}{1.5pt}} }

\newcommand{\botfigrule}{\vspace*{-2pt}%
\noindent{\color{cream}\rule[\figrulesep]{\columnwidth}{1.5pt}} }

\newcommand{\dblfigrule}{\vspace*{-1pt}%
\noindent{\color{cream}\rule[-\figrulesep]{\textwidth}{1.5pt}} }

\makeatother

\twocolumn[
  \begin{@twocolumnfalse}
{\includegraphics[height=30pt]{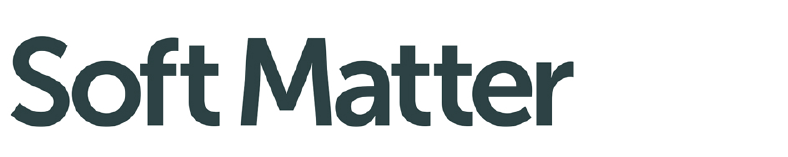}\hfill\raisebox{0pt}[0pt][0pt]{\includegraphics[height=55pt]{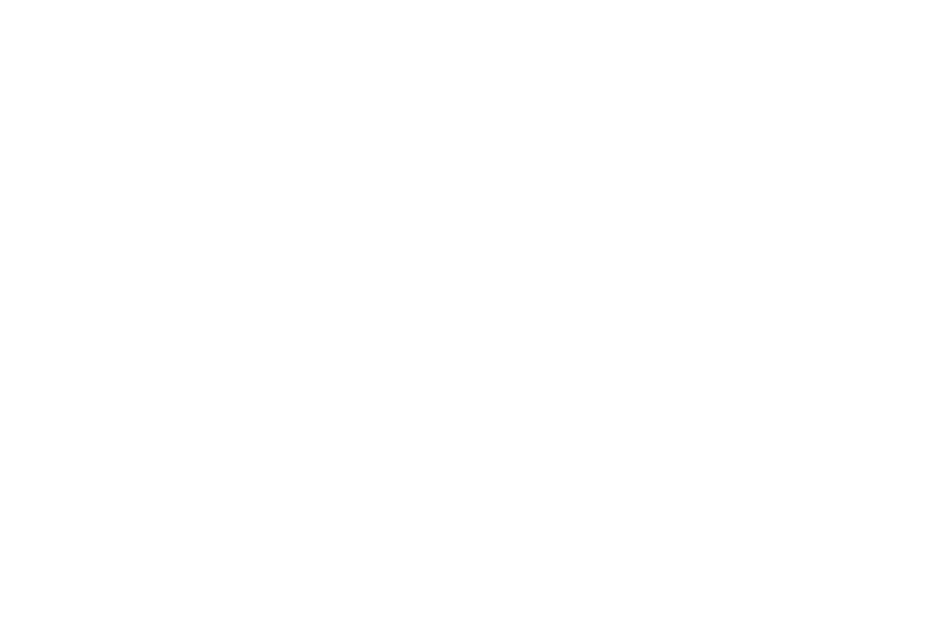}}\\[1ex]
\includegraphics[width=18.5cm]{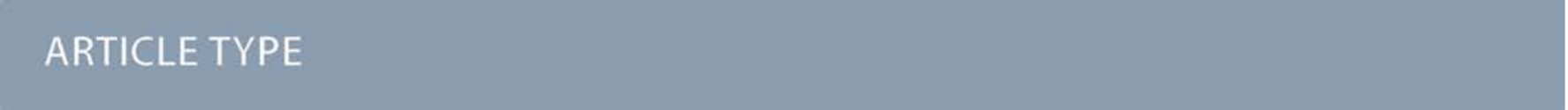}}\par
\vspace{1em}
\sffamily
\begin{tabular}{m{4.5cm} p{13.5cm} }

\includegraphics{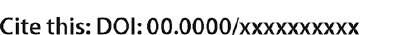} & \noindent\LARGE{\textbf{Mechanochemical enzymes and protein machines as  hydrodynamic force dipoles: The active dimer model}} \\
\vspace{0.3cm} & \vspace{0.3cm} \\

 & \noindent\large{
 Yuto Hosaka,\textit{$^{a}$} 
 Shigeyuki Komura\textit{$^{a\ast}$} and 
 Alexander S. Mikhailov\textit{$^{bc}$}}
 \\

\includegraphics{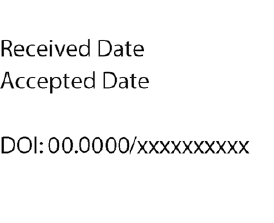} & \noindent\normalsize{
Mechanochemically active enzymes change their shapes within every turnover cycle. Therefore, they induce circulating flows in the solvent around them and behave as oscillating hydrodynamic force dipoles. Because of non-equilibrium fluctuating flows collectively generated by the enzymes, mixing in the solution and diffusion of passive particles within it are expected to get enhanced. Here, we investigate the intensity and statistical properties of such force dipoles in the minimal active dimer model of a mechanochemical enzyme. In the framework of this model, novel estimates for hydrodynamic collective  effects in solution and in lipid bilayers under rapid rotational diffusion are derived, and available experimental and computational data is examined.
} \\

\end{tabular}

 \end{@twocolumnfalse} \vspace{0.6cm}

  ]

\renewcommand*\rmdefault{bch}\normalfont\upshape
\rmfamily
\section*{}
\vspace{-1cm}


\footnotetext{\textit{$^{a}$~Department of Chemistry, Graduate School of Science, Tokyo Metropolitan University, Tokyo 192-0397, Japan}}
\footnotetext{\textit{$^{\ast}$~
E-mail: komura@tmu.ac.jp}}
\footnotetext{\textit{$^{b}$~Computational Molecular Biophysics, WPI Nano Life Science Institute, Kanazawa University, Kakuma-machi, Kanazawa, 920-1192, Japan
}}
\footnotetext{\textit{$^{c}$~
Department of Physical Chemistry, Fritz Haber Institute of the Max Planck Society, Faradayweg 4-6, 14195 Berlin, Germany
}}



\section{Introduction}

Ligand-induced mechanochemical motions are typical for enzymes. Binding or dissociation of a ligand (i.e., substrate or product) to such proteins, as well as chemical reactions within the ligand-bound state, are often accompanied by conformational transitions in them.  Thus, these macromolecules would repeatedly change their shapes in each next turnover cycle. The primary role of mechanochemical motions is to enable and facilitate catalytic reaction events. In the enzymes that operate as protein machines or molecular motors and catalytically convert ATP or GTP, such motions are moreover employed to bring about the required machine function or to generate work.

Since enzymes are in solution, their active conformational changes are accompanied by flows in the fluid around them. Such non-equilibrium flows can affect internal mechanical motions in the enzymes and also influence translational and rotational diffusion of such proteins, as demonstrated by MD simulations for a model protein\cite{cressman} and adenylate kinase.\cite{echeverria}
It has been discussed whether hydrodynamic self-propulsion of enzymes could furthermore occur, in the models where either instantaneous transitions\cite{ajdari,ajdari1} or ligand-induced  continuous conformational motions take place.\cite{iima,sakaue,wolynes}

Lipid bilayers, forming biological membranes, behave as two-dimensional (2D) fluids on submicrometer scales.\cite{Diamant,mj}
Biomembranes often include many active protein inclusions, such as ion pumps or transporters. Essentially, they represent protein machines powered by ATP hydrolysis or other catalytic reactions in them. Within each operation cycle, the shapes of their membrane domains typically change, inducing 2D fluid flows in the lipid bilayer around them.\cite{mj1}
As a result, active protein inclusions might even propel themselves through biomembranes.\cite{mj2}

Collective conformational activity of enzymes and protein machines leads to the development of non-thermal 
fluctuating flows in solution or a lipid bilayer. 
Other particles (i.e., passive tracers) are advected by these non-equilibrium flows, and, as previously shown,\cite{kapralPNAS} increased mixing in such systems and diffusion enhancement should therefore arise.
Additionally, chemotaxis-like effects in the presence of spatial gradients in the concentration or the activity of enzymes can take place.\cite{kapralPNAS}
Remarkably, such phenomena persist even if mechanochemical motions are reciprocal; they do not rely on the presence of self-propulsion for proteins, which is predicted to be weak.\cite{iima,sakaue,wolynes}

Following the original publication,\cite{kapralPNAS} extensive further research has been performed.\cite{PhysD,koyano,PhysRevE,Hosaka17,komura,Yasuda17,Hosaka20,dennison}
The effects of rotational diffusion and of possible nematic ordering for enzymes were considered,\cite{koyano} the phenomena in biomembranes were extensively analyzed,\cite{PhysRevE,Hosaka17} and the theory was extended to viscoelastic media as well.\cite{komura, Yasuda17}
Recently, it was shown that viscosity in dilute solutions of mechanochemically active enzymes should become also reduced.\cite{Hosaka20}
Multiparticle numerical simulations of active oscillatory colloids, explicitly including hydrodynamic effects, were furthermore undertaken and  principal theoretical predictions could thus be verified.\cite{dennison}

At low Reynolds numbers, the flow distribution produced by an object, changing the shape due to internal forces within it, can be characterized in the far field as that corresponding to a hydrodynamical force dipole. If the time-dependent stochastic force dipole of an enzyme is known, the collective hydrodynamic effects in solution of such enzymes are predicted by the mean-field theory.\cite{kapralPNAS}
The difficulty, however, is that  experimental measurements and precise theoretical estimates for intensities and statistical properties of the force dipoles corresponding to actual enzymes are not available yet.
Lacking this knowledge, only rough quantitative estimates for the considered collective hydrodynamic effects could be made so far.

Our present study has two aims and, respectively, it includes two parts.
Section~\ref{sec:stat} corresponds to the first part. Here, the active dimer model is formulated. The active dimer represents a minimal model where ligand-induced mechanochemical motions are reproduced.\cite{kogler,kapralPNAS,dennison,illien2017}
After presenting the model, we undertake an approximate analytical investigation of statistical properties of the force dipoles corresponding to active dimers in subsection~\ref{sec:analytical}, followed by a numerical study in subsection~\ref{sec:numerical}. Quantitative estimates for the intensity of hydrodynamical force dipoles in real enzymes are obtained in subsection~\ref{sec:enzymes}.

Section~\ref{sec:diffusion} corresponds to the second part. Based on the active dimer results, we obtain in subsection~\ref{sec:hydrodynamic} more precise analytical and numerical estimates for the maximal diffusion enhancement for passive particles in solutions of active enzymes, taking into account fast rotational diffusion of enzymes. 
Similar estimates for diffusion enhancement of passive particles in lipid bilayers are derived in subsection~\ref{sec:membrane}.

The results are discussed in Section~\ref{sec:discussion}.
There, we analyze the available experimental and computational data for diffusion enhancement in, respectively, subsections~\ref{sec:experiment} and \ref{sec:computational}.
Conclusions and an outline for the perspectives of further research are provided in Section~\ref{sec:conclusions}.

\section{Statistical properties of force dipoles}
\label{sec:stat}

\subsection{The active dimer model}
\label{sec:dimer}

The simplest mechanical system that gives rise to a hydrodynamical force dipole is a dimer. It consists of two beads 1 and 2 interacting via a potential $u(r)$ that depends on the distance $r=|\boldsymbol{r}_1-\boldsymbol{r}_2|$ between them. The forces acting on the particles are $\boldsymbol{f}_1 = - \partial u/ \partial \boldsymbol{r}_1 =\boldsymbol{f}$ and $\boldsymbol{f}_2 = -\boldsymbol{f}.$ If the dimer is immersed into a viscous fluid, the velocity $\boldsymbol{V}$ of the hydrodynamic flow far enough from the dimer is approximately given by\cite{kapralPNAS}
\begin{align}
V_{\alpha } =\frac{\partial G_{\alpha \beta }}{\partial R_{\gamma }}e_{\beta }e_{\gamma }m,
\end{align}
where $G_{\alpha \beta}(\boldsymbol{R})$ is the mobility tensor depending on the position $\boldsymbol{R}$ of the dimer with respect to the observation point, $\boldsymbol{e} =(\boldsymbol{r}_1 -\boldsymbol{r}_2)/r$ is the unit orientation vector of the dimer, and $m =fr$ is the magnitude of the force dipole. Summation over repeated indices is assumed. The force dipole is present only if there are non-vanishing net interaction forces, i.e., if the distance between the particles in a dimer continues to change.
As in the study,\cite{kapralPNAS} we assume that the Oseen approximation holds. For a dimer, it is justified if the distance between the beads is much larger than their size.

\begin{figure}[t]
\centering
\includegraphics[width=.4\textwidth]{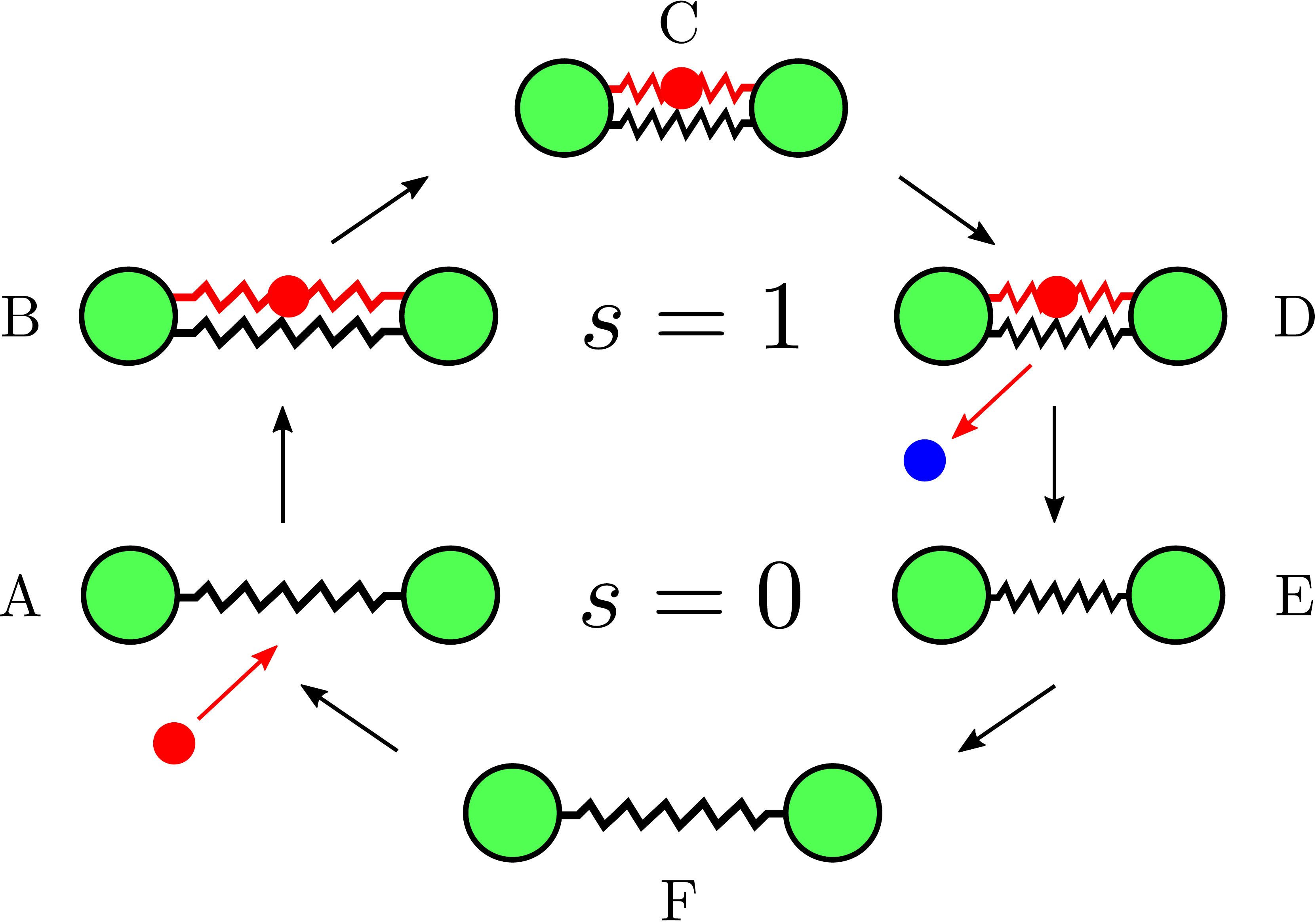}
\caption{The turnover cycle and mechanochemical motions in the active dimer model of an enzyme (see the text).}
\label{fig1}
\end{figure}

The minimal active dimer model has been proposed\cite{kogler,kapralPNAS} (see also review\cite{flechsig}) to imitate mechanochemical conformational motions accompanying a catalytic turnover cycle in an enzyme. 
Note that the dimer model, with non-reactive dissociation of substrate additionally included, was also considered in the study.\cite{illien2017}

The operation mechanism is illustrated in Fig.~\ref{fig1}.
Two identical beads (green) of radius $a$ are connected by an elastic link with a certain natural spring length $\ell_0$  and stiffness $k_0$. A substrate particle (red) arrives (A) and binds as a ligand to the dimer by forming an additional elastic link with stiffness $\kappa$ that connects the two beads (B). The natural length $\ell_{\rm c}$ of this additional link is taken to be shorter than $\ell_0$. Therefore, it tends to contract the dimer until a new equilibrium conformation (C) with a certain distance $\ell_1$ between the beads is reached. Once this has taken place, a chemical reaction, that converts the ligand from the substrate to the product, occurs and the product (blue) is instantaneously released (D). Following the product release, the dimer is in the state E with the spring length $\ell_1$ that is shorter than the natural length $\ell_0$. Therefore, the spring expands and the domains move apart until the equilibrium state (F) is approached again. After that, a new substrate can bind, repeating the turnover cycle.

It is assumed that products are immediately evacuated and therefore we do not consider reverse product binding events. Moreover, possible dissociation events for the substrate are neglected assuming that its affinity is high. 
Note that, since the product is immediately released once it has been formed, the ligand inside our model enzyme is always only in the substrate form.  Therefore, the dimer can be either in the ligand-free ($s=0$) or the ligand-bound ($s=1$) states.

The elastic energies in these two states are
\begin{align}
E_0(x) =\frac{k_0}{2} (x -\ell_0)^2, 
\end{align}
and
\begin{align}
E_1(x) =\frac{k_0}{2} (x -\ell_0)^2 +\frac{\kappa}{2} (x -\ell_{\rm c})^2
=A+\frac{k_1}2 (x -\ell_1)^2,
\end{align}
where $x$ is the distance between the beads and
\begin{align}
&A =\frac{\kappa k_0}{2(k_0+\kappa)}(\ell_0-\ell_{\rm c})^2,~~~~~
k_1=k_0+\kappa, 
\nonumber \\
& \ell_1=\frac{k_0 \ell_0 +\kappa \ell_{\rm c}}{k_0 +\kappa}.
\end{align}

The overdamped dynamics of the dimer in the ligand state $s$ is described by the Langevin equation
\begin{align}
\frac{dx}{dt} = -\gamma \frac{\partial E_s}{\partial x} +\xi(t) , 
\label{eq:Langevin}
\end{align}  
where $\gamma$ is the mobility coefficient.
To account for thermal fluctuations, this equation includes thermal noise, 
\begin{align} 
\langle\xi(t_1) \xi(t_2)\rangle=2\gamma k_{\rm B} T \delta (t_1-t_2) , \label{Eq:FDT}
\end{align}  
where $k_{\rm B}$ is the Boltzmann constant and $T$ is the temperature.

In eqn~(\ref{eq:Langevin}), we have omitted hydrodynamic interactions between the beads.
They were taken into account in the study\cite{illien2017} of diffusion enhancement for a single dimer itself. In the Oseen approximation, such interaction terms are proportional to the small parameter $a/\ell_0$, leading to corrections of the same order for the force dipoles, neglected by us.

Stochastic transitions between the two ligand states take place at constant rates $v_0$ and $v_1$ within narrow windows of width $\rho$ near $x=\ell_0$ and $x=\ell_1$. If probability distributions $p_s(x,t)$ are introduced, they obey a system of two coupled Fokker-Planck equations 
\begin{align}
\frac{ \partial p_0}{ \partial t} &=\frac{\partial}{\partial x}[\gamma k_0(x-\ell_0)p_0] +\gamma k_{\rm B} T\frac{\partial^2p_0}{ \partial x^2} 
\nonumber \\
&+u_1(x)p_1(x)-u_0(x)p_0(x),
\label{p0}
\end{align}
and
\begin{align}
\frac{ \partial p_1}{ \partial t} &=\frac{\partial}{\partial x}[\gamma k_1(x-\ell_1)p_1] +\gamma k_{\rm B} T\frac{ \partial^2p_1}{\partial x^2} \nonumber \\
&+u_0(x)p_0(x)-u_1(x)p_1(x),
\label{p1}
\end{align}
where $u_0(x)=v_0$ for $\ell_0-\rho<x <\ell_0 +\rho$ and vanishes outside of this interval; $u_1(x)=v_1$ for $\ell_1-\rho<x <\ell_1 +\rho$ and zero outside the interval. Note that the rate $v_0$ of substrate binding is proportional to the substrate concentration.

If the transition windows are very narrow, i.e., $\rho \ll \ell_0$ and $\rho \ll \ell_1$, one can use the approximation  
\begin{align}
u_0(x)=\nu_0 \delta(x-\ell_0),~~~~~u_1(x)=\nu_1 \delta(x-\ell_1),
\label{delta}
\end{align}
where $\nu_0=2v_0\rho $ and $\nu_1=2v_1\rho$.

\begin{figure}[t]
\centering
\includegraphics[width=.4\textwidth]{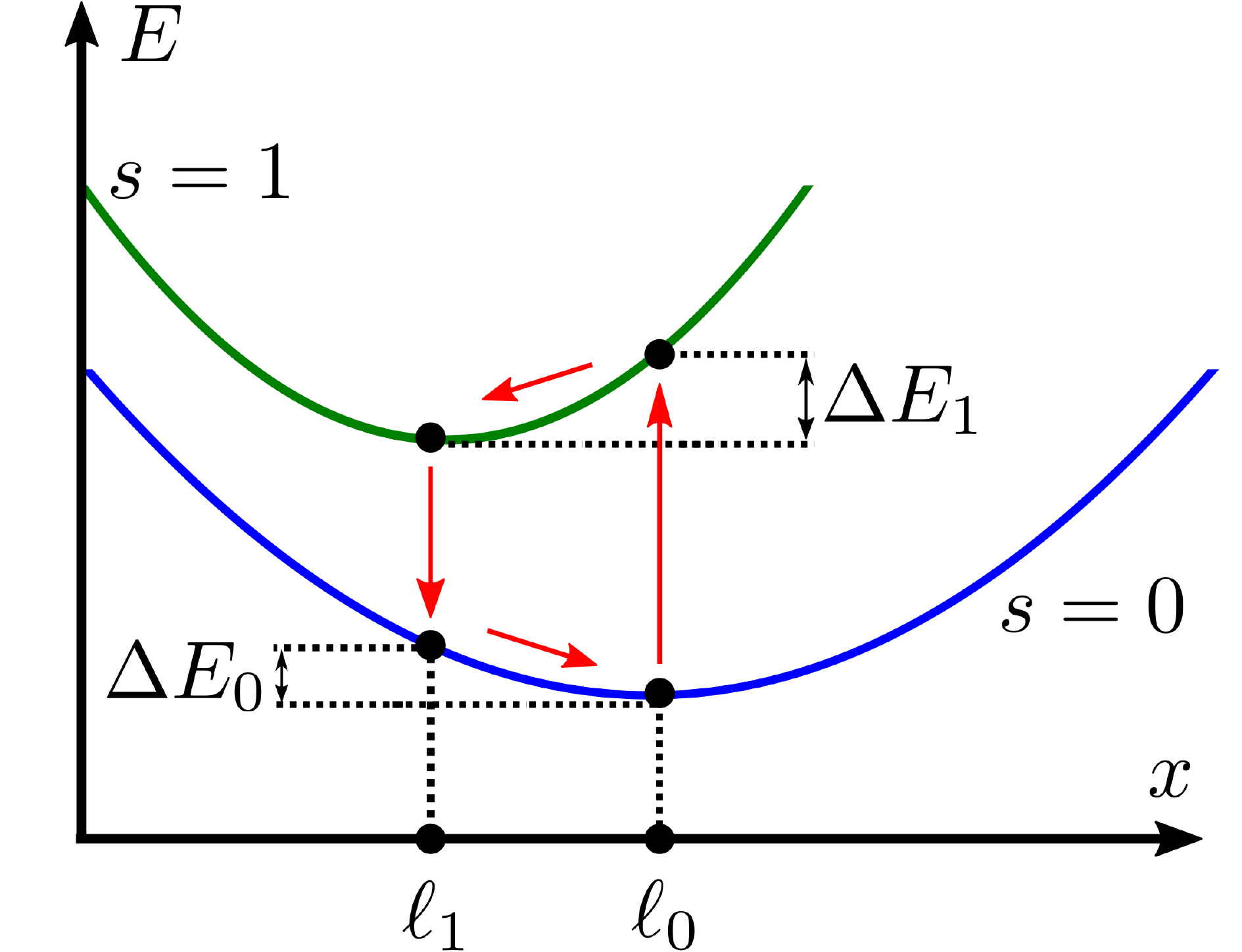}
\caption{The energy diagram of the active dimer.}
\label{fig2}
\end{figure}

Fig.~\ref{fig2} shows the energy diagram of the model. 
Within each cycle, the dimer dissipates in mechanochemical motions the energy $\Delta E =\Delta E_0 +\Delta E_1$ which is furthermore equal to the difference $E_{\rm sub}- E_{\rm prod}$ of the energy $E_{\rm sub}=E_1(\ell_0)-E_0(\ell_0)$ supplied with the substrate and the energy $E_{\rm prod}=E_1(\ell_1)-E_0(\ell_1)$ removed with the product.
We have
\begin{align}
\Delta E=\frac{1}{2}(k_0+k_1)(\ell_0-\ell_1)^2.
\label{supply}
\end{align}
The energy difference $\Delta E$ is always positive and, hence, the considered active dimer represents an exothermic enzyme.

The force dipole of the active dimer is $m = k_0(\ell_0-x) x$ for $s=0$ and $m = k_1(\ell_1-x) x$ for $s=1$. Note that therefore $m \leq k_0 \ell_0^2/4$ for $s=0$ and $m \leq k_1 \ell_1^2/4$ for $s=1$.

When the transition windows are narrow, the probability rate $w_0$ that substrate binding, i.e., a transition to state $s=1$, occurs per unit time in the state $s=0$ is approximately 
\begin{align}\label{w0}
w_0=\nu_0 \sqrt {\frac{k_0}{2\pi k_{\rm B}T }}.
\end{align}
On the other hand, the probability rate $w_1$ that product release, i.e., a transition to state $s=0$, occurs per unit time in the state $s=1$ is then approximately given by
\begin{align}\label{w1}
w_1=\nu_1 \sqrt {\frac{k_1}{2\pi k_{\rm B}T }}.
\end{align}
These equations are derived in Appendix~\ref{app:transition}.
Moreover, the characteristic relaxation times of the dimer in the states $s=0$ and $s=1$ are, respectively, $\tau_0=(\gamma k_0)^{-1}$ and $\tau_1=(\gamma k_1)^{-1}$.

The parameter combinations $w_0\tau_0$ and $w_1\tau_1$ play an important role in determining the kinetic regimes. If the condition $w_0\tau_0 \ll1$ is satisfied, equilibration to thermal distribution in the state $s=0$ usually takes place before a transition to the state $s=1$, i.e., binding of a substrate, occurs. If the opposite condition $w_0\tau_0 \gg1$ holds, such transition takes place immediately after the transition window at $x=\ell_0$ is reached. If $w_1\tau_1 \ll1$, the equilibration takes place in the state $s=1$ before a transition to the state $s=0$, i.e., the reaction and the product release, occurs. In the opposite limit with $w_1\tau_1 \gg1$, the reaction takes place and product becomes released immediately once the respective window at $x=\ell_1$ is reached.

Note that, because the rate $w_0$ is proportional to substrate concentration, the condition $w_0\tau_0 \gg1$ corresponds to the substrate saturation regime for the considered model enzyme. The condition $w_1\tau_1 \ll1$ implies that the enzyme waits a long time before the product is released.

\subsection{Approximate analytical results for force dipoles}
\label{sec:analytical}

At thermal equilibrium in the absence of substrate, $p_1(x) =0$ and 
\begin{align}
p_0(x)=\sqrt {\frac{k_0}{2\pi k_{\rm B}T }}\exp \left[-\frac{k_0}{2k_{\rm B}T}(x -\ell_0)^2\right].
\label{distribution}
\end{align}
Since $m =k_0(\ell_0-x)x$, one can easily find the equilibrium statistical distribution for force dipoles by using the condition $P_{\rm eq}(m)dm=p_0(x)dx$.
Using, for convenience, the dimensionless force dipole magnitude $\widetilde{m}=m/(k_0 \ell_0^2)$ and dimensionless temperature $\theta=k_{\rm B}T/(k_0\ell_0^2)$, we get
\begin{align}
P_{\rm eq}(\widetilde{m}) &=\frac{1}{\sqrt{2\pi(1-4\widetilde{m})\theta}} \biggl\{ \exp\left[ -\frac{1}{8\theta}(1 +\sqrt{1 -4\widetilde{m}})^2 \right] \nonumber \\
&+\exp\left[-\frac{1}{8\theta}(1 -\sqrt{1 -4\widetilde{m}})^2\right]\biggl\}.
\label{dist:m}
\end{align}
If $\theta \ll 1$, this distribution is approximately Gaussian and localized at $m=0$, i.e., 
\begin{align}
P_{\rm eq}(\widetilde{m}) &= \frac{1}{\sqrt{2\pi\theta}}\exp \left(-\frac{\widetilde{m}^2}{2\theta}\right). 
\end{align}

Using the distribution in eqn~(\ref{dist:m}), one finds that the mean force dipole is
\begin{align}
\langle m \rangle_{\rm eq} = -k_{\rm B}T. 
\end{align}
The  correlation function $C(t)=\langle \Delta m(t) \Delta m(0)\rangle $ for variations $\Delta m=m-\langle m \rangle$ of force dipoles is\cite{dennison}
\begin{align}
C_{\rm eq}(t) =k_0 \ell_0^2k_{\rm B}T e^{-|t|/\tau_0} +2(k_{\rm B}T)^2e^{-2|t|/\tau_0},
\label{eq:corr}
\end{align}
where $\tau_0=(\gamma k_0)^{-1}$ is the characteristic relaxation time for the dimer in the state $s=0$.
As shown in Appendix~\ref{app:m}, the exact relation $\langle m \rangle = -k_{\rm B}T$ holds for the dimer in any steady state and, therefore, in any of these limits.

For an active dimer, approximate analytical estimates can be obtained in the four characteristic limits described below.
The two of them (A and C) correspond to low substrate concentrations, with rare turnover cycles controlled by the substrate supply. In regime B, mechanochemical motions are limiting the overall catalytic rate.
In other words, product formation and its release occur once an appropriate conformation ($x=\ell_1$) has been reached. In regime D, the overall kinetic rate is, on the other hand, limited by the waiting time for product formation and release.

\subsubsection*{A.~The limit of $w_0\tau_0\ll1$ and $w_1\tau_1\ll1$}
If these conditions are satisfied, binding of the substrate and product release have large waiting times.
In this limit, there are two almost independent equilibrium subpopulations of dimers in the states $s=0$ and $s=1$.
The relative weights of the subpopulations are $w_1/(w_1+w_0)$ and $w_0/(w_1+w_0)$. Therefore, all statistical properties are given by the sums of contributions from different states taken with the respective weights. Particularly, the correlation function of force dipoles is 
\begin{align}
C(t) &=\frac{w_1}{w_0 +w_1}\Big [k_0 \ell_0^2k_{\rm B}T e^{-|t|/\tau_0} +2(k_{\rm B}T)^2e^{-2|t|/\tau_0}\Big ] \nonumber \\
&+\frac{w_0}{w_0 +w_1}\Big [k_1 \ell_1^2k_{\rm B}T e^{-|t|/\tau_1} +2(k_{\rm B}T)^2e^{-2|t|/\tau_1}\Big ].
\label{correlations}
\end{align}

We can use the above equation to determine the non-equilibrium part of the fluctuation intensity of force dipoles 
\begin{align}
\langle \Delta m^2 \rangle_{\rm A}=\langle \Delta m^2 \rangle-\langle \Delta m^2 \rangle_{\rm eq}.
\label{eq:activemsquare}
\end{align}
Because $\langle \Delta m^2 \rangle = C(0)$, we have
\begin{align}
\langle \Delta m^2 \rangle_{\rm A} =\frac{w_0}{w_0 +w_1}\left(k_1\ell_1^2-k_0\ell_0^2\right)k_{\rm B}T.
\label{fluctuations}
\end{align}
As follows from eqn~(\ref{eq:corr}), the equilibrium fluctuation intensity is 
\begin{align}
\langle \Delta m^2 \rangle_{\rm eq}=k_0 \ell_0^2k_{\rm B}T +2(k_{\rm B}T)^2.
\label{equil:intensity}
\end{align} 
Since the effective binding rate $w_0$ of the substrate is proportional to its concentration $c$, i.e., $w_0=\eta c$, eqn~(\ref{fluctuations}) yields the Michaelis-Menten form of the dependence of $\langle \Delta m^2 \rangle_{\rm A}$ on the substrate concentration.

Remarkably, the catalytic activity of the model enzyme can thus lead not only to some enhancement, but also to {\it reduction} of fluctuations of the force dipoles. 
According to eqn~(\ref{fluctuations}), reduction should be observed if $k_1\ell_1^2<k_0\ell_0^2$. Under this condition, the ligand-bound dimer ($s=1$) is characterized by a lower fluctuation intensity of force dipoles than the free dimer ($s=0$).

\subsubsection*{B.~The limit of  $w_0 \tau_0\gg1$ and $w_1\tau_1\gg 1$}
In this limit, transitions take place once the respective transitions windows are entered.
If additionally the conditions $k_0\ell_0^2 \gg k_{\rm B}T$ and $k_1\ell_1^2 \gg k_{\rm B}T$ are satisfied, thermal fluctuations can be neglected and the dimer essentially behaves as a deterministic oscillator. Then, the solution can be obtained by integrating eqn~(\ref{eq:Langevin}) with appropriate boundary conditions. This yields $ x(t) =\ell_1 +(\ell_0 -\ell_1-\rho)e^{-t/\tau_1} $
for $0<t<T_1$ and $ x(t) =\ell_0 +(\ell_1-\ell_0+\rho)e^{-(t -T_1)/\tau_0}$
for $T_1 <t <T_{\rm c}$. Here, $T_{\rm c}$ is the oscillation period of the active dimer and $T_1$ is the duration of the cycle time when the dimer is in the ligand-bound state $s=1$. If transition windows are narrow, i.e., the condition $\rho\ll(\ell_0 -\ell_1)$ is satisfied, we approximately have
\begin{align}
T_1 =\tau_1\ln\left(\frac{\ell_0 -\ell_1}{\rho }\right),
\label{time1}
\end{align}
and
\begin{align}
T_{\rm c} =(\tau_0+\tau_1)\ln \left(\frac{\ell_0 -\ell_1}{\rho }\right).
\label{period}
\end{align}
The respective time-dependent force dipole is $m(t)= k_1(\ell_1-x)x$ for $0<t<T_1$ and $m(t) = k_0(\ell_0-x)x$ for $T_1 <t <T_{\rm c}$. Hence, it is negative for $s=1$ and positive for $s=0$.

The force dipole varies within the interval $m_{\text{min}} < m < m_{\text{max}}$, where the minimum value $m_{\text{min}} = -k_1 \ell_0 (\ell_0-\ell_1)$ is taken at $t =0 ,$ i.e., in the state $s=1$ just after substrate binding, and the maximum value $m_{\text{max}} =k_0 \ell_1 (\ell_0-\ell_1)$ is reached at $t =T_{1}$, in the state $s=0$ just after product release (here we again assume that transition windows are narrow). Note that, if thermal fluctuations were present, the force dipoles could however have also taken the values outside of this interval.

It can be checked by direct integration that the period-averaged force dipole for the deterministic active dimer is $\langle m(t) \rangle_{\text{det}} =0$. The correlation function for the deterministic oscillating dimer is defined as the period average
\begin{align}
C_{\text{det}}(t)=\frac{1}{T_{\rm c}}\int_{0}^{T_{\rm c}}dh\, m(t+h)m(h).
\end{align}
The explicit analytical form of this periodic correlation function is too complicated and we do not give it (analytical results for correlation functions are also omitted below in the limits C and D).

The mean-square intensity of force dipoles is $\langle m(t)^2 \rangle_{\text{det}} = C_{\text{det}}(0)$. In the limit $\rho\rightarrow 0$, we approximately have 
\begin{align}
& \langle m(t)^2 \rangle_{\text{det}} =\frac{k_0k_1}{12(k_0+k_1)}
\left[\ln\left(\frac{\ell_0-\ell_1}{\rho}\right)\right]^{-1}(\ell_0-\ell_1)^2 \nonumber \\
&\times\Big[k_0(\ell_0^2+2\ell_0\ell_1+3\ell_1^2)+k_1(3\ell_0^2+2\ell_0\ell_1+\ell_1^2)\Big].
\label{detfluct}
\end{align}
When $k_1\sim k_0$, this equation yields the scaling $\langle m(t)^2 \rangle_{\rm det}\sim k_0^2$.

\subsubsection*{C.~The limit of $w_0\tau_0\ll1$ and $w_1\tau_1\gg1$}
If these conditions are satisfied, the model enzyme waits a long time for binding of a substrate (because the substrate concentration is low), but then it performs a rapid reaction cycle.
An approximate solution in this regime can be obtained if, additionally, the conditions $k_0\ell_0^2\gg k_{\rm B}T$ and $k_1\ell_1^2\gg k_{\rm B}T$ are satisfied, i.e., that thermal fluctuations are weak. Moreover, we shall assume that the transition window for substrate binding is narrow, i.e., the approximation in eqn~(\ref{delta}) holds for $u_0(x)$.

In this case, the dependence $x(t)$ consists of a sum of statistically independent rare pulses, each corresponding to one reaction cycle:
\begin{align}
x(t)=\sum_jz(t-t_j),
\end{align}
where $z(t)=\ell_1 +(\ell_0 -\ell_1)e^{-t/\tau_1}$ for $0<t<T_1$ and $z(t)=\ell_0+(\ell_1-\ell_0)e^{-(t-T_1)/\tau_0}$ for $t>T_1$, with $T_1$ given by eqn~(\ref{time1}). 
The pulses appear at random time moments $t_j$ and the probability of their appearance per unit time is $w_0$.

Moreover, we also have  
\begin{align}
m(t)=\sum_j\zeta(t-t_j),
\end{align}
where $\zeta(t)= k_1(\ell_1-z(t))z(t)$ for $0<t<T_1$ and $\zeta(t) = k_0(\ell_0-z(t))z(t)$ for $t>T_1$.

Hence, this represents a random Poisson process. Its first two statistical moments are approximately $\langle m(t)\rangle=0$ and 
\begin{align}\label{fluctC}
&\langle m^2(t)\rangle=w_0\int_0^\infty dt\,\zeta^2(t)=\frac{1}{12}w_0\tau_0(\ell_0-\ell_1)^2 \nonumber \\
&\times \left[k_0^2(\ell_0^2+2\ell_0\ell_1+3\ell_1^2)+k_0k_1(3\ell_0^2+2\ell_0\ell_1+\ell_1^2)\right].
\end{align}
Taking into account eqn~(\ref{w0}), we notice that, when $k_1\sim k_0$, the scaling $\langle m^2(t)\rangle\sim k_0^{3/2}$ should hold.

\subsubsection*{D.~The limit of $w_0\tau_0\gg1$ and $w_1\tau_1\ll1$}
This situation corresponds to substrate saturation and a long waiting time for the reaction and product release in the ligand-bound state. 
A derivation, similar to that given above, shows that, if $k_0\ell_0^2\gg k_{\rm B}T$ and $k_1\ell_1^2\gg k_{\rm B}T$, we approximately have $\langle m(t)\rangle=0$ and 
\begin{align}\label{fluctD}
& \langle m^2(t)\rangle =\frac{1}{12}w_1\tau_1(\ell_0-\ell_1)^2 \nonumber \\
&\times \left[k_1^2(3\ell_0^2+2\ell_0\ell_1+\ell_1^2)+k_0k_1(\ell_0^2+2\ell_0\ell_1+3\ell_1^2)\right].
\end{align}
If we take into account eqn~(\ref{w1}), it can be noticed that, when $k_1\sim k_0$, scaling $\langle m^2(t)\rangle\sim k_0^{3/2}$ is again obtained.

\subsection{Numerical simulations}
\label{sec:numerical}

Numerical simulations can yield statistical properties of force dipoles for selected parameter values in the regions where there are no approximate analytical results. Below in this section, we focus on the situation under substrate saturation, but with the waiting time for product release comparable to the conformational relaxation time (thus, lying between the limits B and D).
We shall consider a situation where the condition $k_{\rm B}T \ll k_0\ell_0^2$ is satisfied, so that thermal fluctuations are relatively weak.

Before proceeding to simulations, the model was non-dimensionalized. The dimensionless variables were $\tilde{t}=t/\tau_0$, 
$\tilde{x}=x/\ell_0$, and $\widetilde{m}=m/(k_0\ell_0^2)$. The dimensionless transition rates were 
$\tilde{v}_0=v_0\tau_0$ and $\tilde{v}_1=v_1\tau_0$, while the dimensionless temperature was $\theta=k_{\rm B}T/(k_0\ell_0^2)$. Stochastic differential equation (\ref{eq:Langevin}) was numerically integrated, complemented by transitions between the ligand states.

In the simulations, we had $\ell_1=0.55 \ell_0$, $k_1=2k_0$, and  $\rho=0.01\ell_0$. We have kept constant $\tilde{v}_1=2$, but varied  the parameter $\tilde{v}_0$. 
A relatively low dimensionless temperature $\theta=0.0018$ was chosen to satisfy the condition $k_{\rm B}T \ll k_0\ell_0^2$.
Under such choice, $\langle m(t)^2 \rangle_{\text{det}}/\langle \Delta m^2 \rangle_{\rm eq}=19.3$ and $w_1\tau_1=0.27$.

Note that, because of the last condition, there was a significant random variation in the waiting times for substrate conversion and product release. Moreover, waiting times for substrate binding, characterized by the rate $w_0$, could also vary. These effects kept the model stochastic even when thermal noise was small.

\begin{figure}[t]
\centering
\includegraphics[width=.3\textwidth]{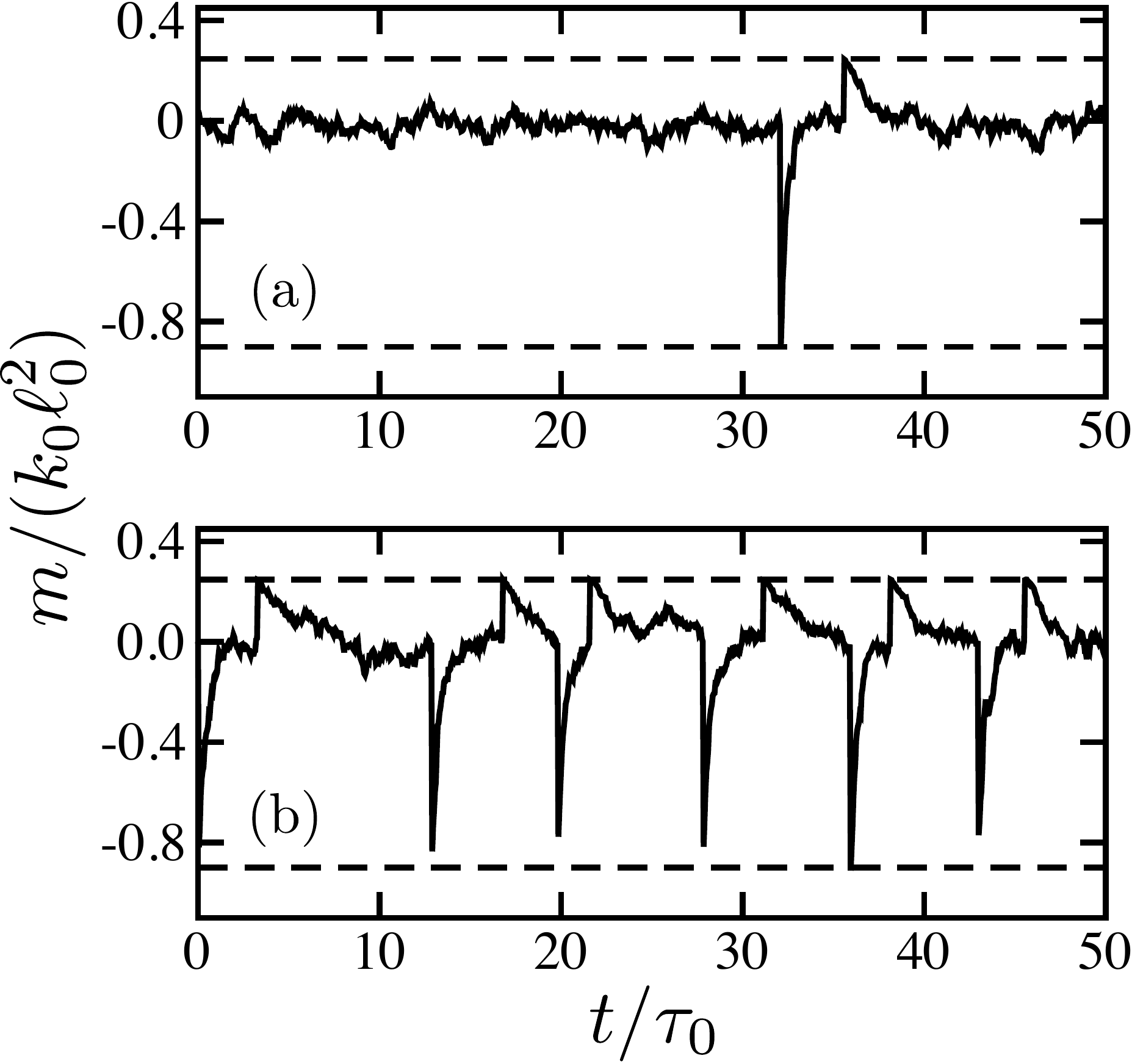}
\caption{Time dependence of dimensionless force dipoles $\widetilde{m}=m/(k_0\ell_0^2)$ on time for (a) $\tilde{v}_0=0.03$ and (b) $\tilde{v}_0=3$. Dashed lines show the lower bound $\widetilde{m}_{\text{min}}=-0.9$ for the deterministic oscillatory dimer and the absolute upper bound $\widetilde{m}_{\rm max}=0.25$ for force dipoles.
} 
\label{fig3}
\end{figure}

Fig.~\ref{fig3} shows typical time dependences of the force dipoles. 
In Fig.~\ref{fig3}(a), the waiting time for substrate binding is long.
Therefore, the dimer spends most of the time in the ligand-free state $s=0$. 
Within the time shown, only one turnover cycle has taken place. For the force dipole, the cycle consists of a negative spike, just after binding of the substrate, and the following positive spike, just after the product release. 
In Fig.~\ref{fig3}(b), the substrate binding rate is increased. As a result, the dimer is frequently cycling, already resembling an oscillator. Nonetheless, the random variation of the times between the cycles is relatively large.

\begin{figure}[t]
\centering
\includegraphics[width=.3\textwidth]{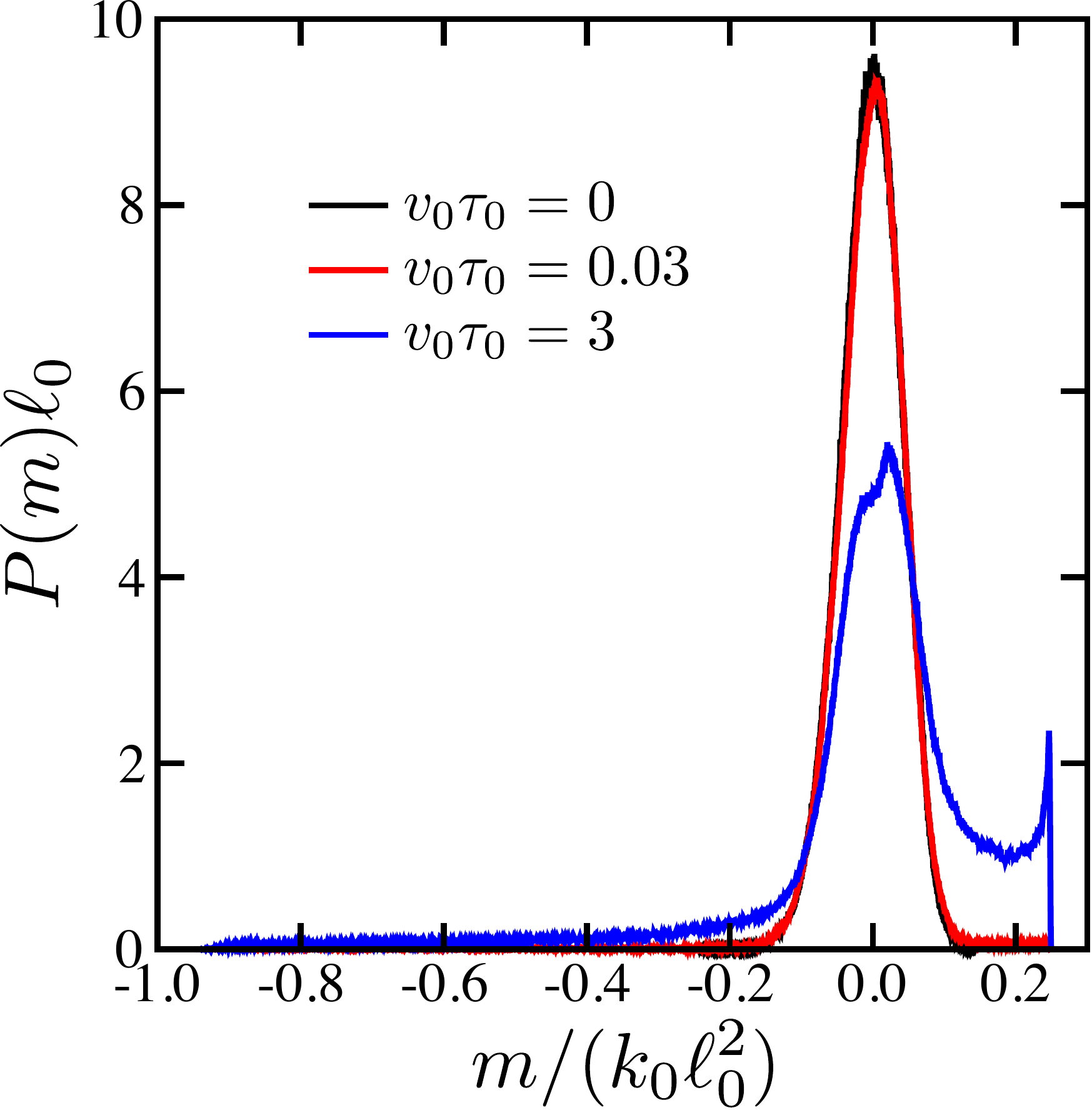}
\caption{Probability distributions of force dipoles $\widetilde{m}$ for passive (black curve, $v_0=0$) and active (red curve, $\tilde{v}_0=0.03$, and blue curve, $\tilde{v}_0=3$) dimers.}
\label{fig4}
\end{figure}

Probability distributions of force dipoles are shown in Fig.~\ref{fig4}.
The black curve is the distribution for passive dimers in the absence of the substrate, given by eqn~(\ref{dist:m}). 
It represents a narrow Gaussian peak at $m=0$. The distribution at $\tilde{v}_0=v_0\tau_0=0.03$ (red) is almost indistinguishable from it. The blue curve is the distribution for active dimers corresponding to Fig.~\ref{fig3}(b).
Now, the distribution is more broad and the central peak is smaller. The tail on the left side from the peak and the shoulder on its right side are due to the non-equilibrium activity of force dipoles.

\begin{figure}[t]
\centering
\includegraphics[width=.3\textwidth]{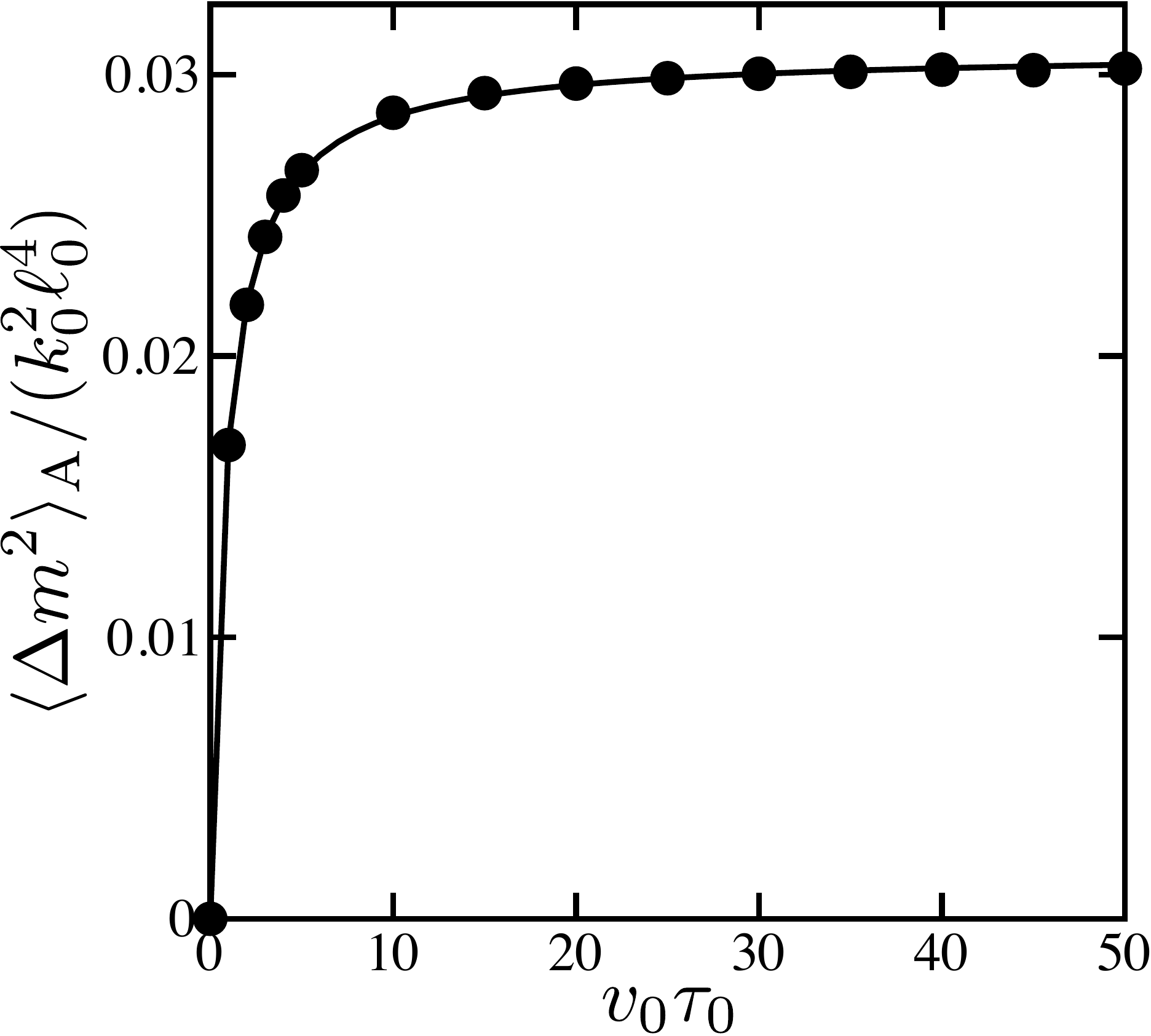}
\caption{Dependence of the non-equilibrium part $\langle \Delta m^2 \rangle_{\rm A}$ of the fluctuation intensity of force dipoles on the substrate binding rate $v_0$ (dots). The solid curve is a fit to the Michaelis-Menten function.}
\label{fig5}
\end{figure}

The dependence of the non-equilibrium part of the fluctuation intensity of force dipoles, eqn~(\ref{eq:activemsquare}), on the substrate binding rate $v_0$, proportional to substrate concentration, is shown in Fig.~\ref{fig5}.
It can be well fitted to the Michaelis-Menten function (the solid curve).
The saturation magnitude is close to the value of 0.033 predicted at such parameters for the deterministic dimer by eqn~(\ref{detfluct}).

\begin{figure}[t]
\centering
\includegraphics[width=.3\textwidth]{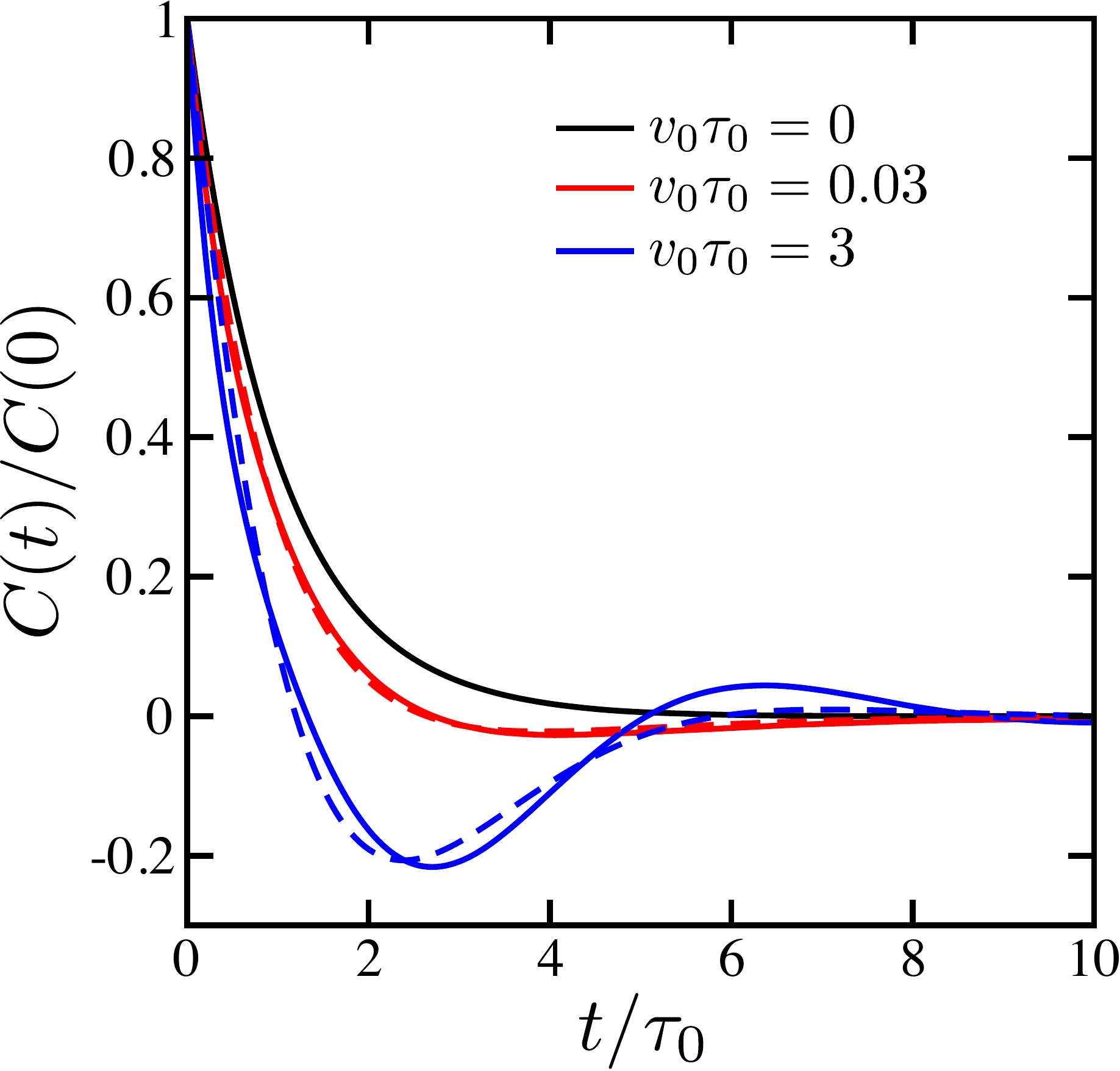}
\caption{Normalized correlation functions of force dipoles at different substrate binding rates: $v_0=0$ (absence of substrate, black), $\tilde{v}_0=0.03$ (red), and $\tilde{v}_0=3$ (blue). The correlation function for passive dimers (black) is given by eqn~(\ref{eq:corr}). Dashed curves are fits to the dependence in eqn~(\ref{damped}).}
\label{fig6}
\end{figure}

Normalized correlation functions of force dipoles at different substrate binding rates are shown in Fig.~\ref{fig6}.
In the absence of the substrate (for $v_0=0$) the dependence is monotonous (it is given by eqn~(\ref{eq:corr})). 
As the substrate concentration is increased, damped oscillations in the correlation function become observed, thus signaling the onset of the active oscillatory behavior that prevails over the thermal noise.

\begin{figure}[t]
\centering
\includegraphics[width=.3\textwidth]{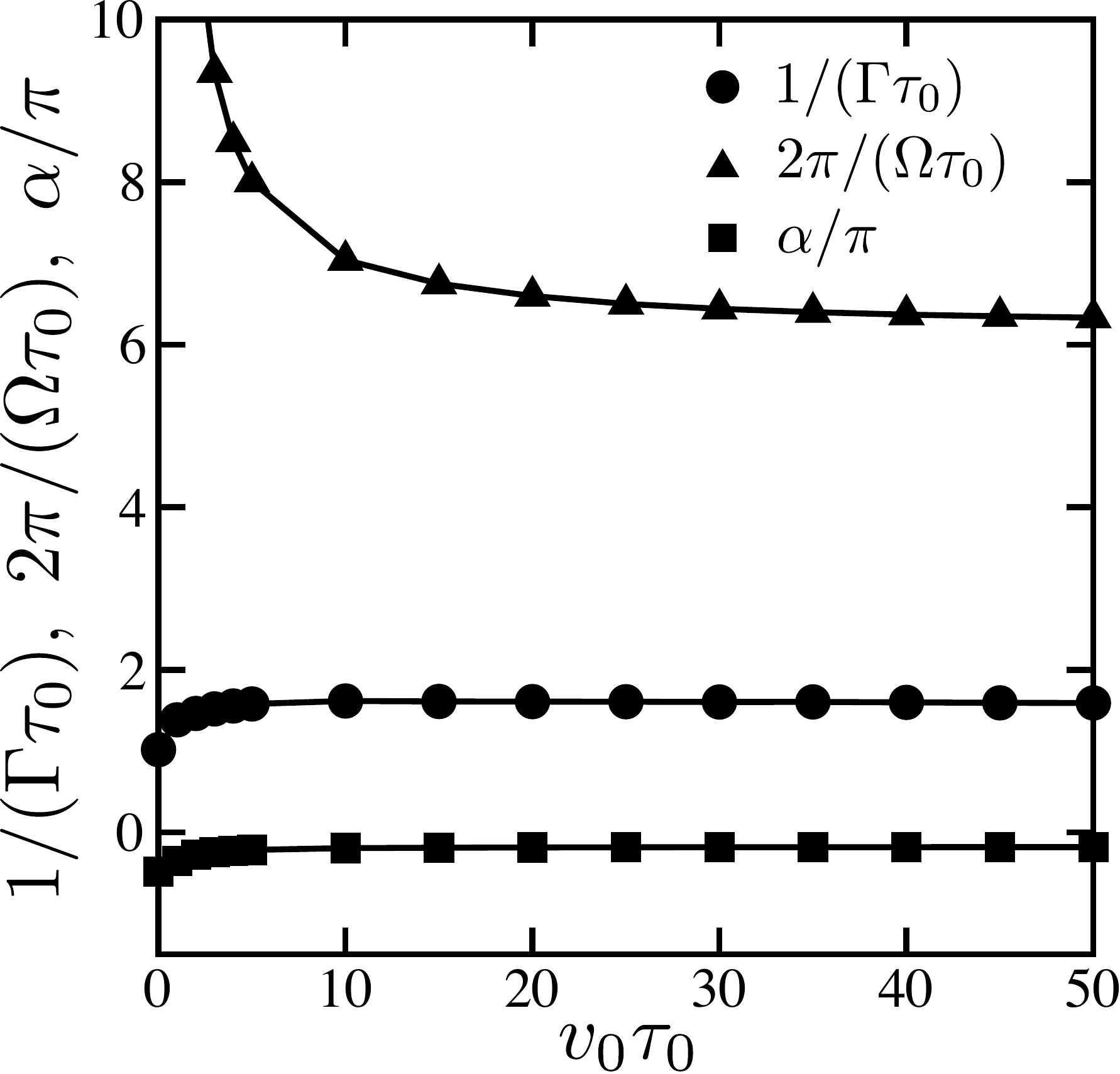}
\caption{The dependences of the relaxation time $1/\Gamma$ (circles), oscillation period $2\pi /\Omega$ (triangles) and phase shift $\alpha$ (squares) on substrate binding rate $v_0$.}
\label{fig7}
\end{figure}

The correlation functions could be fitted (dashed curves in Fig.~\ref{fig6}) to the dependence
\begin{align}
C(t)/C(0)=\frac{1}{\cos\alpha}\exp(-\Gamma |t|)\cos(\Omega |t|-\alpha).
\label{damped}
\end{align} 
Fig.~\ref{fig7} shows how the dimensionless relaxation time $1/(\Gamma \tau_0)$, the dimensionless oscillation period $2\pi /(\Omega\tau_0)$ and the phase shift $\alpha$ depend on the substrate binding rate. The oscillation period under saturation conditions is still larger than $T_{\rm c}/\tau_0=5.7$ for the deterministic dimer according to eqn~(\ref{period}).
This is because of an additional waiting time for product release. The characteristic relaxation time is about $1/(\Gamma\tau_0)=2$.

It should be stressed that the form in eqn~(\ref{damped}) of the correlation function would not hold in the deterministic limit. 
Indeed, the oscillations stay harmonic in the limit of an infinite correlation time. However, the deterministic oscillations are actually non-harmonic, as seen in Fig.~\ref{fig3}.

There are two effects that make the dimer model stochastic, i.e., the thermal noise in the dynamical equation (\ref{eq:Langevin}) and random transitions between the ligand states $s=0$ and $s=1$. When $\theta \rightarrow 0$, the thermal noise vanishes, but random transitions between the states nonetheless remain. 
This second stochastic effect is responsible for the decay in the correlation function. As shown in Appendix~\ref{app:FP}, the dependence of the correlation function in eqn~(\ref{damped}) corresponds to an approximate solution of the master equations (\ref{p0}) and (\ref{p1}).

\subsection{Estimates for hydrodynamic force dipoles of enzymes}
\label{sec:enzymes}

Above, statistical properties of force dipoles were analyzed in the framework of an idealized model of the active dimer. Now, the obtained results can be applied to approximately estimate the force dipoles for real enzymes and protein machines. To do this, the relationship between such a simple model and the actual proteins needs to be first discussed.

Proteins fold into a definite conformation that however incorporates many different substates.
Slow dynamics of proteins represents wandering over a Markov network of such metastable conformational substates.\cite{noe}
In all-atom molecular dynamics (MD) simulations, transitions within tens of nanoseconds to the nearest metastable states can be clearly seen. Long MD simulations show motions over a set of these states extending to the millisecond timescales.\cite{shaw} 
Single-molecule fluorescence correlation spectroscopy experiments with cholesterol oxidase revealed that thermal conformational fluctuations in this enzyme, in the absence of the substrate, had correlations persisting even over about 1.5\,s time.\cite{xie,lerch} 
In the coarse-grained structure-based simulations of proteins, such as modeling based on elastic networks, the rugged atomic energy landscape becomes smoothed,\cite{bahar} thus yielding continuous slow conformational dynamics described by a set of effective collective coordinates.

In a detailed study of adenylate kinase,\cite{kern} combining all-atom MD simulations with single-molecule fluorescence resonance energy transfer and NMR, it was found that, in this characteristic mechanochemical enzyme, conformational substates lie along a trajectory that connects the initial open apo conformation to the final catalytically efficient closed state. Thus, the energy landscape has a valley that guides towards the optimal protein state; the motion along such a valley can be described by a single coordinate. Similar organization of the energy landscape has been noticed in  structure-based coarse-grained modeling of protein machines and molecular motors, such as myosin V and $\text{F}_1$-ATPase\cite{togashi} and HCV helicase.\cite{flechsig1}

Typically, mechanochemical enzymes and molecular machines represent proteins with domain structure. Slow functional conformational dynamics in these proteins consists in relative motions of the domains that can be often characterized by a single coordinate, such as a hinge angle or a distance between the centers of mass of two protein domains. This leads to the reduced models for  proteins, with just one or a few mechanical coordinates.\cite{flechsig}
The active dimer is a model belonging to this class. Note that previously a similar simple model with three beads was employed to estimate the magnitude of self-propulsion effects in the enzymes.\cite{wolynes}
In the framework of the active dimer model, statistical properties of force dipoles in different kinetic regimes can be analyzed and characteristic order-of-magnitude estimates for the intensity of such dipoles for typical enzymes and protein machines can be derived.

In Section~\ref{sec:analytical}, four kinetic regimes have been outlined. The two of them (A and C) correspond to low substrate concentrations, with rare turnover cycles controlled by the substrate supply. 
Below, we focus on the substrate saturation regimes B and D where high catalytic activity and, thus, the strongest non-equilibrium force dipoles can be expected.

In regime B, mechanochemical motions are limiting the overall catalytic rate.
In other words, product formation and its release occur once an appropriate conformation ($x=\ell_1$) has been reached. Such regime is characteristic, for example, for adenylate kinase where the turnover time is limited by the time (about 1\,ms) of the conformational transition from the open to the closed state, with the reaction ${\rm AMP+ATP\rightarrow2ADP}$ rapidly occurring once the latter state is reached.\cite{kern}

In regime D, the overall kinetic rate is, on the other hand, limited by the waiting time for product formation and release. This regime is typical for protein machines and motors such as myosin V. In each operation cycle of this molecular motor, catalytic hydrolysis of substrate ATP into product ADP takes place. The cycle duration of 66\,ms under ATP saturation is limited by waiting for ADP release.\cite{thirumalai}
The conformational transition from the open to the closed state, i.e., the lever-arm swing after ATP binding, takes place within a much shorter millisecond time.

The principal parameters of the active dimer model are stiffness constants $k_0$ and $k_1$ and inter-domain distances $\ell_0$ and $\ell_1$ in the open ($s=0$) and closed ($s=1$) conformations, respectively.
The typical size of a protein is of the order of tens of nanometers and this would be also the characteristic distances $\ell_0$ and $\ell_1$ between the domains. Moreover, if the open and closed states are distinctly different, as, for example, in adenylate kinase or myosin, the change $\Delta\ell=\ell_0-\ell_1$ is comparable in magnitude to $\ell_0$ and $\ell_1$. As characteristic values for order-of-magnitude estimates, one can, for example, choose $\ell_0=10$\,nm and $\ell_1=5$\,nm in the open and the closed states, respectively.

In the active dimer, two domains (beads) are connected by a spring. In real proteins, they can be, instead, connected by a hinge with the elastic energy
\begin{align}
    E=\frac{1}{2}K(\Theta-\Theta_0)^2,
\end{align}
which depends on the deviation of the hinge angle $\Theta$ from the equilibrium angle $\Theta_0$. This can also be approximately written as
\begin{align}
    E=\frac{1}{2}k(x-\ell_0)^2,
\end{align}
so that the hinge is described as an elastic spring with $x=\ell\Theta$,  $\ell_0=\ell\Theta_0$ and the effective stiffness $k=K/\ell^2$, where $\ell$ is the characteristic linear size of the domains connected by the hinge.

The stiffness of the converter hinge in myosin V was estimated in single-molecule experiments by Kinoshita with coworkers\cite{kinoshita} to be about $K=5$\,$k_{\rm B}T/\text{rad}^{2}$ both in the open and the closed states. 
On the other hand, the data of high-speed AFM observations by Ando with coworkers\cite{ando} corresponds to a higher value of $K=23$\,$k_{\rm B}T/\text{rad}^{2}$. The difference may be due to the fact that the hinge becomes softer for larger angles. In our estimates below, we take $K=10$\,$k_{\rm B}T/\text{rad}^{2}$. 
Choosing $\ell=10$\,nm, this leads to $k= 0.1$\,$k_{\rm B}T/\text{nm}^2$.

As noticed above, in adenylate kinase, the overall turnover rate under substrate saturation in an enzyme is limited by conformational transitions between the open and closed states (and hence the turnover rate is about $10^3$\,$\text{s}^{-1}$).
The maximum intensity of force dipoles can be estimated by using eqn~(\ref{detfluct}).
If the parameter values $k_0=k_1=0.1$\,$k_{\rm B}T/\text{nm}^2$, $\ell_0=10$\,nm, $\ell_1=5$\,nm and $\rho=1$\,nm are chosen, the non-equilibrium mean-square fluctuation intensity $\langle \Delta m^2 \rangle_{\rm A}$ of force dipoles in such enzymes is estimated as approximately 80\,[pN$\cdot$nm]${}^2$.
This is similar to the previous estimate of 100\,[pN$\cdot$nm]${}^2$ in Ref.\cite{kapralPNAS} based on typical stall forces in molecular motors.

In more slow enzymes and protein machines with the turnover numbers of tens per second, the turnover is limited by product formation and its release. In this case, the intensity of force dipoles can be estimated using eqn~(\ref{fluctD}). There, the rate $w_1$ of product formation and release is approximately the same as the overall turnover rate, whereas $\tau_1$ corresponds to the conformational transition time. Choosing the turnover rate of 15\,$\text{s}^{-1}$, as in myosin V, and the conformational transition time of 1\,ms and keeping the same other parameters as above, the non-equilibrium mean-square fluctuation intensity of force dipoles can then be estimated as about 4\,[pN$\cdot$nm]${}^2$.
This is much smaller than the above estimate for fast enzymes because non-thermal mechanical forces are only generated in conformational transitions of about 1\,ms in duration.
It represents only a small fraction of the entire cycle time of tens of milliseconds in such enzymes or protein machines.

While typical enzymes have turnover times between milliseconds and tens of milliseconds, there are also very slow enzymes, such as tryptophan synthase with the turnover time of 0.5\,s,\cite{stryer} and enzymes that are very fast, such as catalase (17\,$\mu$s) or urease (59\,$\mu$s).\cite{bustamante}
Moreover, transition times from open to close confirmations can be also very short in some enzymes. 
For example, for phosphoglycerate kinase (PGK), neutron spin-echo spectroscopy yields conformational transition times of the order of tens of nanoseconds.\cite{inoue}
This was also observed in coarse-grained MD simulations for PGK.\cite{schonfield}
Therefore, it is interesting to discuss under what general conditions stronger force dipoles can be expected in enzymes.

Equation~(\ref{supply}) relates the energy (generally, enthalpy) dissipated in mechanochemical motions within the turnover cycle of an active dimer to the stiffness of the dimer and the magnitude of conformational changes in it. While it has been derived for an idealized model, it can also be used for order-of-magnitude estimates in real enzymes. Taking, for example, $k_0=k_1=0.1$\,$k_{\rm B}T/\text{nm}^2$ and $\ell_0-\ell_1=10$\,nm for myosin, we obtain $\Delta E=10$\,$k_{\rm B} T$, which is in reasonable agreement with the energy of about 20\,$k_{\rm B} T$ supplied to this molecular motor with ATP (only half of this energy is used in the power stroke).

Note that, assuming for simplicity that $k_0=k_1=k$, eqn~(\ref{supply}) can be also written as
\begin{align}\label{stiff}
  k=\frac{\Delta E}{\Delta\ell^2},  
\end{align}
thus expressing the stiffness in terms of the energy $\Delta E$ dissipated in mechanochemical motions and the conformational change $\Delta \ell=\ell_0-\ell_1$. 
An enzyme is stiffer if the same energy is dissipated within a conformational transition of a smaller magnitude.

Suppose that conformational changes are indeed small in an enzyme and, moreover, its turnover rate is limited by conformational transitions within the cycle. Then, eqn~(\ref{detfluct}) can be used to estimate the intensity of force dipoles. For approximate numerical estimates, it can be written in the form 
\begin{align}
\langle \Delta m^2 \rangle= \zeta_0 k^2\ell_0^2(\ell_0-\ell_1)^2,
\label{max}
\end{align}
where $\zeta_0$ is a dimensionless factor of order unity that also includes the logarithmic term and we have taken $k_0=k_1=k$. Note that this estimate holds assuming that the force dipoles in the catalytically active enzyme are much stronger than those due to thermal fluctuations in the absence of substrate.

Substituting $k$ from eqn~(\ref{stiff}), a simple order-of-magnitude estimate is obtained  
\begin{align}\label{intens}
\langle \Delta m^2 \rangle=\zeta_1 \left(\frac{\ell_0}{\Delta \ell}\right)^2\Delta E^2,
\end{align}
where $\zeta_1$ is another dimensionless factor of order unity. Moreover, by using eqn~(\ref{equil:intensity}) and (\ref{stiff}), we furthermore get
\begin{align}
    \frac{\langle \Delta m^2 \rangle}{\langle \Delta m^2 \rangle_\text{eq}}=\zeta_1\frac{\Delta E}{k_{\rm B} T},
\end{align}
if the condition $k\ell_0^2\gg k_{\rm B} T$ holds.

These results show that the intensity of force dipoles is strongly sensitive to the magnitude of mechanochemical motions within the turnover cycle. Moreover, they show that, in strongly exothermic enzymes,  force dipoles are greatly enhanced when catalytic activity takes place.

The above-mentioned catalase and urease enzymes are not only exceptionally fast, but also highly exothermic, with $\Delta H=100$\,${\rm kJ}/\text{mol}$ for catalase and $\Delta H= 59.6$\,${\rm kJ}/\text{mol}$ for urease.\cite{bustamante}
Hence, large energies of 42\,$k_{\rm B} T$ or 25\,$k_{\rm B} T$ are released in them and dissipated into heat within very short microsecond cycle times.
Furthermore, at least for catalase, it is known that functional conformational changes are involved within the turnover cycle, but their magnitude is small.\cite{gouet} 
It has been previously proposed\cite{bustamante} that chemoacoustic intramolecular effects caused by strong heat release may even lead to hydrodynamic self-propulsion of these enzymes, although subsequent examination could not confirm this.\cite{wolynes}
These enzymes do not have a domain structure and therefore the results of our analysis based on the dimer model are not directly applicable to them. Nonetheless, they suggest that hydrodynamic force dipoles in them may be very strong.

\section{Diffusion enhancement for passive particles in solutions in active enzymes}
\label{sec:diffusion}

The most important application of the obtained results for force dipoles is that they allow to obtain more accurate analytical and numerical estimates for diffusion enhancement of passive particle in solutions of active enzymes. 
In the previous studies,\cite{kapralPNAS,koyano} the magnitude of diffusion enhancement has been expressed in terms of the statistical properties of hydrodynamic force dipoles. However, because these properties were only poorly known, precise estimates for such magnitude could not be obtained.
This has led to difficulties in the interpretation of experimental results and in the analysis of the computational data. 
In this section, we use the statistical properties of force dipoles for active dimers, determined above in Section~\ref{sec:stat}, to estimate the diffusion enhancement effects for enzymes in water solution and for active protein inclusions in biomembranes.

\subsection{Diffusion effects of enzymes in water solutions}
\label{sec:hydrodynamic}

As previously shown \cite{kapralPNAS,koyano}, the change $D_{\rm A}$ in the diffusion coefficient of passive \textit{tracer} particles in a three-dimensional (3D) solution is given by\ 
\begin{align}
D_{\rm A}= \frac{n}{60 \pi \mu^2 \ell_\text{cut}}(\chi-\chi_{\rm eq}),
\label{enhancement}
\end{align}
where $n$ is the concentration of active enzymes, $\mu$ is viscosity, and $\ell_\text{cut}$ is a microscopic cut-off length of the order of a protein size.
Moreover, we have
\begin{align}
\chi=\int_0^{\infty}dt\,C(t)\sigma(t),
\label{chi}
\end{align}
where  $C(t)$ is the correlation function of force dipoles corresponding to the enzymes and $\sigma(t)$ is the orientational correlation function for them; $\chi_{\rm eq}$ is given by the same equation, but with $C(t)$ replaced by $C_{\rm eq}(t)$.

The orientational correlation function has the form 
\begin{align}
\sigma(t)=\exp(-t/\tau_{\rm rot}),
\label{rotation}
\end{align}
where $\tau_{\rm rot}$ is the orientational correlation time. 
As seen from eqn~(\ref{enhancement}), (\ref{chi}) and (\ref{rotation}), the magnitude of diffusion enhancement is sensitive to the relationship between the correlation time of force dipole and the orientational correlation time.

According to the Stokes equation, rotational diffusion coefficient for a spherical particle of radius $R$ is 
\begin{align}\label{StokesRot}
D_{\rm rot}=\frac{k_{\rm B}T}{8\pi\mu R^3}.
\end{align}
The orientational correlation time is defined as $\tau_{\rm rot}=1/D_{\rm rot}$. Since proteins are not spheres, their orientational correlation  times are shorter than given by the Stokes estimate. Even in crowded solutions, they do not exceed a microsecond.\cite{hummer,spoel,feig}
For active dimers, the orientational correlation times can be estimated by using eqn~(\ref{StokesRot}) with $R=\ell_0$.

For the active dimer model, the correlation functions of force dipoles and, therefore, their correlation times were analytically determined at equilibrium [see eqn~(\ref{eq:corr})] and in the limit A [see eqn~(\ref{correlations})].
Moreover, they were also numerically determined, as shown in Fig.~\ref{fig6}.
As follows from these results, the correlation times for force dipoles are determined by conformational relaxation times $\tau_0$ and $\tau_1$. In the discussion of conformational relaxation phenomena in proteins in subsection~\ref{sec:enzymes}, we have noticed that slow conformational relaxation processes, involving relative domain motions in real enzymes, would usually lie in the microsecond to millisecond range. Hence, correlation times for force dipoles of the enzymes would be typically longer than their orientational correlation times.

Below, we assume that the orientational correlation time is much shorter than the correlation time for force dipoles. By using eqn~(\ref{rotation}) and putting $C(t)\approx C(0)=\langle \Delta m^2 \rangle$ in eqn~(\ref{chi}), we approximately find
\begin{align}\label{deltachi}
  \chi-\chi_{\rm eq}=\tau_{\rm rot} \langle \Delta m^2 \rangle_{\rm A}.  
\end{align}
Hence, the change in the diffusion coefficient can be estimated as
\begin{align}
D_{\rm A}= \frac{\tau_{\rm rot}n}{60 \pi \mu^2 \ell_{\rm cut}}\langle \Delta m^2 \rangle_{\rm A}.
\label{DA}
\end{align}
Substituting approximate analytical expressions for $\langle \Delta m^2 \rangle_{\rm A}$ in different limiting regimes, obtained in subsection~\ref{sec:analytical}, or using the numerical data from subsection~\ref{sec:numerical}, we can now estimate and analyze diffusion enhancement.

Particularly, it was found in subsection~\ref{sec:numerical} that $\langle \Delta m^2 \rangle_{\rm A}$ for active dimers has a Michaelis-Menten dependence on the substrate concentration. As follows from the above equation, the same dependence should hold for $D_{\rm A}$. As could have been expected, the highest diffusion enhancement is reached under the condition of substrate saturation for the enzymes. 
Under substrate saturation, the intensity of force dipoles depends, as demonstrated in subsections~\ref{sec:analytical} and \ref{sec:numerical}, on the properties of the turnover cycle of an enzyme. The highest intensity is found in regime B, i.e., when there is no long waiting time for product release, and the turnover rates are limited by conformational transitions within the catalytic cycle (adenylate kinase is an example of such an enzyme).

In such asymptotic regime, $\langle \Delta m^2 \rangle_{\rm A}$ is given by eqn~(\ref{intens}) provided that the condition $k\ell_0^2 \gg k_{\rm B}T$ holds. Substituting this expression into eqn~(\ref{DA}), we obtain
\begin{align}
\frac{D_{\rm A}}{D_{\rm T}}=\frac{\nu R_0}{R_0+\ell_0}\left(\frac{\ell_0}{\Delta\ell}\right)^2\left(\frac{\ell_0}{\ell}\right)^3\left(\frac{\Delta E}{k_{\rm B}T}\right)^2.
\label{ratio}
\end{align}
Here, $R_0$ is the radius of a tracer particle, $\ell_0$ is the characteristic size of an enzyme (i.e., the dimer length), $\Delta\ell$ specifies the magnitude of the conformational change, $\Delta E$ is the free energy supplied with the substrate and dissipated within each cycle. For the equilibrium diffusion constant $D_{\rm T}$, the Stokes equation
\begin{align}
D_{\rm T}=\frac{k_{\rm B}T}{6\pi\mu R_0}. 
\label{StokesTrans}
\end{align}
has been used. Moreover, the microscopic cut-off length is chosen as $\ell_{\text{cut}}=\ell_0+R_0$ and $\nu=4\pi\zeta_1/5$ is a numerical factor of order unity.

Equation~(\ref{ratio}) can be employed to estimate the maximum relative diffusion enhancement for passive particles that can be obtained, under substrate saturation, in water solutions of enzymes. For numerical order-of-magnitude estimates, we consider exothermic enzymes with $\Delta E=10$\,$k_{\rm B}T$ and 
$\Delta \ell=0.1\ell_0$. As the enzyme concentration, we take $n=1$\,$\mu$M. This corresponds to a non-crowded solution where the mean distance between the enzymes is about ten times larger than their size ($\ell\sim10\ell_0$).
Moreover, we consider passive particles with the sizes comparable to that of an enzyme ($R_0\sim\ell_0)$.  Under these conditions, we have $D_{\rm A}\sim 10 D_{\rm T}$, i.e., diffusion of tracer particles is ten times faster in the solution of catalytically active enzymes.

Dependence of diffusion enhancement on the orientational correlation time is additionally discussed in Appendix~\ref{app:DE}.

\subsection{Diffusion effects of active protein inclusions in biomembranes}
\label{sec:membrane}

It is known that, on the length scales shorter than the Saffman-Delbr\"{u}ck length of about a micrometer, lipid bilayers behave as 2D fluids.\cite{Diamant}
Similar to enzymes in water solutions, active protein inclusions (such as ion pumps or transporters) can cyclically change their shapes inside a lipid bilayer within each ligand turnover cycle. 
Hence, they behave as hydrodynamical force dipoles within a fluid lipid bilayer. 
Therefore, diffusion enhancement is expected for biomembranes when non-equilibrium conformational activity of proteins takes place.\cite{kapralPNAS}

A significant difference to water solution is that, for the biomembranes as 2D fluids, hydrodynamic diffusion enhancement effects are non-local. For such systems, eqn~(\ref{enhancement}) is replaced by\cite{kapralPNAS,koyano}
\begin{align}
  D_{{\rm A},\alpha\alpha'}(\mathbf{R})=\frac{1}{32\pi^2\mu_{\rm 2D}^2}(\chi-\chi_{\rm eq})
 \int d\mathbf{r}\, \frac{r_\alpha r_{\alpha'}}{r^4}n_{\rm 2D}(\mathbf{R}+\mathbf{r}).
\end{align}
Here, $\chi$ is again given by eqn~(\ref{chi}) with $\sigma(t)$ being the planar orientational correlation function for protein inclusions. Moreover, $\mu_{\rm 2D}$ is the 2D viscosity of the lipid bilayer, related as $\mu_{\rm 2D}=h\mu_{\rm 3D}$ to its 3D viscosity $\mu_{\rm 3D}$ (where $h$ is the bilayer thickness); $n_{\rm 2D}$ is the 2D concentration of active inclusions within the membrane.

For numerical estimates, we assume that active proteins occupy a small circular region (a raft) of radius $R_{\rm m}$ (shorter than the Saffman-Delbr\"{u}ck length) within a membrane. 
Then, diffusion enhancement for a passive particle of radius $R_0$ located in the center of the disc is\cite{kapralPNAS,koyano}
\begin{align}
D_{\rm A}=\zeta_{\rm m}\frac{n_{\rm 2D}}{\mu_{\rm 2D}^2}(\chi-\chi_{\rm eq}),
\label{raft}
\end{align}
where $\zeta_{\rm m}=(1/32\pi)\ln({R_{\rm m}/\ell_{\rm cut})}$, $\ell_{\rm cut}=R_0+\ell_0$, and $\chi$ is given by the integral in eqn~(\ref{chi}) where, however, $\sigma(t)$ is the planar orientational correlation function for proteins inside a membrane.

The viscosity $\mu_{\rm 3D}$ of lipid bilayers is about 10$^3$ times higher than that of water and, therefore, both translational and rotational diffusion is much slower in them. From experiments, it is known that diffusion constants for proteins in lipid bilayers are about $D_{\rm T}=10^{-10}$\,${\rm cm}^2/{\rm s}$, i.e., about 10$^3$ times smaller than in water for similar proteins. One can therefore expect that rotational diffusion of proteins in lipid bilayers would be slowed by about a factor of 10$^3$ too, yielding orientational correlation times $\tau_{\rm rot}$ that might approach a millisecond, still being shorter than the turnover time of an enzyme.

The magnitude of diffusion enhancement in eqn~(\ref{raft}) can be determined by modeling protein inclusions as active dimers that lie flat in the membrane. Then, the same estimate (\ref{intens}) for $\langle \Delta m^2 \rangle_{\rm A}$ can be used. Combining all terms, diffusion enhancement in eqn~(\ref{raft}) for a passive particle in the center of a protein raft
approximately is
\begin{align}
D_{\rm A}=\nu_{\rm m}\tau_{\rm rot}\left(\frac{\ell_0}{\Delta\ell}\right)^2\left(\frac{\Delta E}{h\ell_{\rm 2D}\mu_{\rm 3D}}\right)^2,
\end{align}
where the dimensionless prefactor is $\nu_{\rm m}=\zeta_1\zeta_{\rm m}$ and $\ell_{\rm 2D}=n_{\rm 2D}^{-1/2}$ is the mean distance between inclusions in the membrane.

To obtain a characteristic order-of-magnitude estimate, the 3D viscosity of the lipid bilayer is chosen as $\mu_{\rm 3D}=1$\,Pa$\cdot$s and the thickness of the bilayer as $h=1$\,nm. For protein inclusions, we assume that $\Delta E=10$\,$k_{\rm B}T$ and $\Delta \ell\sim\ell_0$. The orientational correlation time is taken to be $\tau_{\rm rot}=100$\,$\mu$s and the mean lateral distance between the proteins is $\ell_{\rm 2D}=10$\,nm. For such parameter values, the maximal possible diffusion enhancement under substrate saturation conditions is about  $D_{\rm A}=10^{-9}$\,${\rm cm}^2/{\rm s}$. 
For comparison, Brownian diffusion constants for proteins in lipid bilayers are of the order of $10^{-10}$\,${\rm cm}^2/{\rm s}$ and diffusion constants for lipids are about $10^{-8}$\,${\rm cm}^2/{\rm s}$.

\section{Discussion}
\label{sec:discussion}

Using the results of our study, available experimental and computational data on diffusion enhancement in solutions of catalytically active enzymes can be discussed.

\subsection{Experimental data}
\label{sec:experiment}

Diffusion enhancement for the enzymes has been reported in solutions of several catalytically active enzymes, at the concentrations varying between 1\,nM and 10\,nM.\cite{golestanian,dey,granick,Xu,bustamante} With the exception of aldolase\cite{golestanian} (for which, however, the enhancement could not be independently confirmed\cite{Hess}), all these enzymes were exothermic and had high turnover rates of about $10^4\,{\rm s}^{-1}$. The enhancement was reported not only for the enzymes themselves, but also for inert molecules (tracers) surrounding them.\cite{dey, granick}
The enzyme concentration dependence of the diffusion enhancement effects could \textit{not} however be detected.\cite{Xu}

It does not seem plausible that such experimental data can be understood in the framework of the original theory\cite{kapralPNAS} and its subsequent extensions, including the present work. The fact that a significant diffusion enhancement (by tens of percent) was observed already at low nanomole concentrations can still be perhaps explained by assuming that, for some reasons, the force dipoles of specific enzymes with high catalytic turnover rates were exceptionally strong. However, the absence of a dependence of the experimentally observed diffusion enhancement on the enzyme concentration clearly contradicts the theory\cite{kapralPNAS} where diffusion enhancement arises as a collective hydrodynamic effect. Effectively, diffusion enhancement was observed in the experiments\cite{golestanian,dey,granick,Xu,bustamante} already for single molecules of enzymes.

When our study was completed, an interesting publication\cite{wang2020} has appeared where diffusion enhancement was demonstrated by a different method for several other reactions. Since catalyst's diffusion was not affected by its concentration, this was again a single-particle property not covered by the theory.\cite{kapralPNAS}

Experiments on optical tracking of particles in animal cells\cite{weitz} and in bacteria or yeast\cite{parry} have been furthermore performed. They have shown that, when metabolism was suppressed (by depletion of ATP), diffusion dropped to undetectable levels\cite{weitz,fodor} or it was much slowed down and replaced by subdiffusion characteristic for a colloidal glass.\cite{parry}
Strong reduction of diffusion under metabolism suppression was moreover found in various cytoplasm extracts.\cite{mizuno}

It should be also noted that diffusion enhancement has been experimentally observed within chromatin in a living biological cell.\cite{weber}
This was explained by active operation of molecular machines involved in transcription and translation of DNA.\cite{grosberg}

The cytoplasm of a living cell represents a crowded solution of proteins. In bacteria, the volume fraction of proteins in cytosol is about 30 percent,\cite{FEMS} with the highest concentrations of the order of 100\,$\mu$M reached for glycolysis enzymes. Most of the enzymes in the cell are mechanochemical, i.e., they exhibit conformational changes in their catalytic cycles. Typical turnover times of enzymes in a biological cell are of the order of 10\,ms.

According to the previous\cite{kapralPNAS} and current estimates, substantial diffusion enhancement due to hydrodynamic collective effects should thus be expected under metabolism in the cytoplasm. There are, however, also other mechanisms that can contribute to diffusion enhancement in the cells.

The cytoskeleton of animal cells represents an active gel, with numerous myosin molecular motors operating within it. It is known that the activity of the motors can lead to development of non-equilibrium fluctuations in the cytoskeleton which induce in turn fluctuations and diffusion enhancement in the cytosol.\cite{levine,fodor,komura}
The skeleton of bacteria and yeast is however passive; moreover, metabolic diffusion enhancement in such cells could also be observed when their skeleton was chemically resolved.\cite{parry} Therefore, the active gel mechanism\cite{levine} cannot account for the effects observed in them.

On the other hand, under high crowding characteristic for cytoplasm, proteins are frequently colliding and direct interactions between them often take place.\cite{hummer,feig}
It is known that, for dense colloids, glass behavior can be expected, with the transport and relaxation phenomena strongly slowed down in them.\cite{hunter}
Indeed, such behavior could be observed both in the cells\cite{parry} and in the extracts\cite{mizuno} in the absence of metabolism.

It has been recently shown that, when the particles forming a glass-like colloid, cyclically change their shapes, the colloid gets fluidized and classical transport properties become restored.\cite{EPL,ikeda}
Even in the absence of hydrodynamic interactions, conformational activity of proteins, at the rates of energy supply of about $10$\,$k_{\rm B}T$ per a protein molecule per a cycle, can lead to diffusion enhancement by one order of magnitude.\cite{EPL} This provides an additional, non-hydrodynamic, mechanism that can contribute to the experimentally observed diffusion enhancement in living biological cells.

In summary, the analysis of the available experimental data reveals that the predicted diffusion enhancement\cite{kapralPNAS} for passive particles caused by collective catalytic activity of enzymes could not so far been reliably confirmed.

\subsection{Computational data}
\label{sec:computational}

Large-scale computer simulations for colloids of active dimers have been performed by Dennison, Kapral and Stark.\cite{dennison}  In these simulations, the solvent was explicitly included and the multiparticle collision dynamics (MPCD) approximation\cite{kapralReview} was employed, thus allowing to fully account for hydrodynamic effects.

To facilitate the comparison, we first give a summary of the essential parameter values in the study,\cite{dennison} using the current notations employed by us. The natural lengths of the dimer in two ligand states were $\ell_0$ and $\ell_1=\ell_0/2$, and the spring constants were $k_0$ and $k_1=2k_0$. The dimensionless spring constant $k\ell_0^2/(k_{\rm B}T)$, characterizing stiffness of the dimer, was varied between 144 and 1440. The energy $\Delta E=(1/2)(k_0+k_1)(\ell_0-\ell_1)^2$, supplied to a dimer and dissipated by it as heat within a single cycle, was changing therefore between 121.5\,$k_{\rm B}T$ and 1215\,$k_{\rm B}T$. 
The simulations were performed under substrate saturation conditions. Product formation and release were possible within a window of half-width $\rho=0.025\ell_0$ near $x=\ell_1$. The rate $v_1$ of this transition could be varied in the simulations by a factor of 5.

The Langevin equation (\ref{eq:Langevin}) with viscous friction and thermal noise was not used. Instead, collisions between the two beads of the dimer and the solvent particles were explicitly taken into account in the framework of MPCD. 
For a single passive dimer, the equilibrium correlation function of force dipoles $C_{\rm eq}(t)$ was computed yielding the correlation time for fluctuations of its force dipole; this function could be well fitted to the theoretical dependence in eqn~(\ref{eq:corr}).
Note that, when $k_0\ell_0^2/(k_{\rm B}T)\gg1$, the relaxation time $\tau_0=(\gamma k_0)^{-1}$ of the dimer should be close to this correlation time. Moreover, we have $\tau_1=(\gamma k_1)^{-1}=\tau_0/2$. Using such estimates, it can be shown that $w_1\tau_1$ varied between 0.001 and 0.1 in the simulations.\cite{dennison}
Because substrate saturation was assumed, conditions $w_0\tau_0\gg1$ and $w_1\tau_1\ll1$ corresponding to the limit D in Section~\ref{sec:analytical} were therefore approximately satisfied.

For single active dimers, correlation functions $C(t)$ of force dipoles were determined.\cite{dennison}
They showed damped oscillations and were similar to the correlation function for $v_0\tau_0=3$ in Fig.~\ref{fig6}.
The correlation times varied, but remained of the same order of magnitude as the correlation time of the passive dimer. The force-dipole intensity $\langle \Delta m^2 \rangle$ of active dimers was by about an order of magnitude larger than $\langle \Delta m^2 \rangle_{\rm eq}$ for the passive ones. Depending on the parameters, it scaled as $k_0^\alpha$ with the exponent $\alpha$ in the range between 1.2 and 1.6, comparable with the exponent of 1.5 in eqn~(\ref{fluctD}).

Orientational correlation functions $\sigma(t)$ were furthermore computed for single dimers.\cite{dennison}
Remarkably, it was found that the orientational correlation time $\tau_{\rm rot}$  was sensitive to the conformational activity of the dimer, getting shorter by about an order of magnitude when such activity was switched on. Nonetheless, in all simulations $\tau_{\rm rot}$ was larger than the force dipole correlation time.

Multiparticle 3D computer simulations of colloids formed by active dimers were further performed.\cite{dennison}
In the simulations, the truncated potential
\begin{align}
u(r)=4\epsilon\left[\left(\frac{2r_0}{r}\right)^{48}-\left(\frac{2r_0}{r}\right)^{24}+\frac{1}{4}\right],
\end{align}
for $r<2^{1/24}(2r_0)$ and zero otherwise, with $\epsilon=2.5$\,$k_{\rm B}T$ and $r_0=1.075\ell_0$, was used to describe steep repulsive interactions between the beads belonging to different dimers. The interaction radius $r_0$ was chosen as defining the radius of a bead.

Since distances $\ell_0$ and $\ell_1=0.5\ell_0$ in the open and closed dimer conformations were both smaller than $2r_0=2.15\ell_0$, large overlaps between the beads in a dimer were present in the simulations. However, this did not affect the internal dimer dynamics because there were no repulsive interactions between the beads in the same dimer. Additionally, the simulated system included one passive tracer particle of radius $0.5\ell_0$.

The volume fraction $\phi$ occupied by dimers was determined by taking into account the overlaps, but assuming that all dimers were in the equilibrium open state with the length of $\ell_0$. Because, under substrate saturation conditions, they were however mainly found in the closed state with an even stronger overlap, such definition overestimated the actual volume fraction by a factor of up to two.

Due to the crowding effects, diffusion of a passive particle in the system of inactive dimers decreased with the volume fraction of them. The diffusion reduction at the highest taken volume fraction $\phi=0.266$ was less than ten percent, indicating that this colloidal system was still far from the glass transition threshold.\cite{hunter}

When the dimers were active, diffusion of tracers was increasing instead with the dimer volume fraction $\phi$. For the most stiff active dimers with $k_0\ell_0^2/(k_{\rm B}T)=1440$ and the kinetic regime with $w_1\tau_1$ about 0.1, relative diffusion enhancement of $D_{\rm A}/D_{\rm T}=0.3$ could be observed\cite{dennison}
at the dimer volume fraction of $\phi=0.266$. For the least stiff dimers with $k_0\ell_0^2/(k_{\rm B}T)=144$, diffusion enhancement by 5 percent was seen at $\phi=0.133$.

Thus, collective hydrodynamic effects of active enzymes on diffusion of passive particles could be computationally confirmed. To speed up the calculations, model enzymes in the study\cite{dennison} were chosen however to be unusually rapid (with the turnover times shorter than the rotational diffusion time) and unusually exothermic (with the heat release of hundreds of $k_{\rm B}T$ per a turnover cycle). It would be therefore important to undertake such simulations also for the parameters closer approaching those of the real enzymes.

\section{Conclusions and outlook}
\label{sec:conclusions}

To our knowledge, the present work is the first study where hydrodynamic force dipoles of mechanochemical enzymes have been systematically analyzed. Although the analysis has been performed for an idealized model, order-of-magnitude estimates for the intensity of such dipoles for characteristic enzymes, such as adenylate kinase, and for protein machines, such as myosin, have been obtained.

We have also examined for what kinds of enzymes strong hydrodynamic effects may be expected. Our analysis reveals that, in the framework of the investigated model, these should be very rapid, highly exothermic and stiff enzymes, where the energy is dissipated in mechanical motions of a small amplitude. It is interesting to note that these general conditions are indeed satisfied, for example, for catalase or urease.

Using the derived statistical properties of force dipoles in the dimer model, more accurate estimates for diffusion enhancement for surrounding passive particles in solutions of active enzymes were obtained.

Based on these results, currently available experimental and computational data has been examined. We have concluded that, while the collective hydrodynamic effects of diffusion enhancement have been principally confirmed in the computational study,\cite{dennison} further work is needed to bring simulations closer to the parameter region corresponding to real enzymes.

On the experimental side, we have concluded that the data on diffusion enhancement in weak nanomole solutions of several fast exothermic enzymes cannot be explained in the framework of the theory\cite{kapralPNAS} and alternative explanations for them should be sought. In experimental studies of diffusion phenomena in living cells and in cellular extracts, additional work is needed to distinguish possible hydrodynamic contributions from the effects of direct collisions between active proteins and the resulting kinetic crowding effects. Large-scale numerical simulations of crowded active colloids including hydrodynamic interactions between the particles are to be performed. It should be also pointed out that, although the effects of diffusion enhancement are also predicted for biomembranes crowded with active protein inclusions, experiments and numerical multiparticle simulations of such phenomena are still missing today; it would be interesting to carry them out.

It was not the aim of the present work to provide a review of all proposed mechanisms for diffusion enhancement effects. Especially, we have not considered possible origins of diffusion enhancement for single catalytically active enzymes, even though this question attracts much attention in view of the recent research.\cite{Xu, granick, wang2020}
Our focus was on diffusion enhancement for passive particles caused by hydrodynamic collective stirring of the solution by a population of active particles that cyclically change their shapes, but do not propel themselves.

In the future, the active dimer model can be used to develop stochastic thermodynamics of mechanochemical enzymes. It would be important to investigate in detail hydrodynamic effects, accompanying functional conformational transitions, in all-atom or coarse-grained molecular dynamics simulations for specific enzymes and protein machines.

\section*{Conflicts of interest}
There are no conflicts to declare.

\section*{Acknowledgements}

Stimulating discussions with K.\ K.\ Dey, R.\ Kapral, H.\ Kitahata and Y.\ Koyano are gratefully acknowledged.
Y.H.\ acknowledges support by a Grant-in-Aid for JSPS Fellows (Grant No.\ 19J20271) from the Japan Society for the Promotion of Science (JSPS).
Y.H.\ also thanks for the hospitality of the Fritz Haber Institute of the Max Planck Society, where part of this research was conducted in the support of the scholarship from TMU.
S.K.\ acknowledges support by a Grant-in-Aid for Scientific Research (C) (Grant No.\ 18K03567) from the JSPS, and support by a Grant-in-Aid for Scientific Research on Innovative Areas ``Information Physics of Living Matters'' (Grant No.\ 20H05538) from the Ministry of Education, Culture, Sports, Science and Technology of Japan. 
S.K.\ and A.S.M.\ acknowledge the support by Grant-in-Aid for Scientific Research (C) (Grant No.\ 19K03765) from the JSPS.

\appendix
\section{Transition probabilities}
\label{app:transition}

When transitions between the states $s=0$ and $s=1$ are rare, the solution of the master equations (\ref{p0}) and (\ref{p1}) can be approximately sought in the form
\begin{align}
p_s(x,t)=\pi_s(t) p(x,t|s),
\end{align}
where $\pi_s(t)$ is the probability to find the dimer in the ligand state $s$ and $p(x,t|s)$ is the probability distribution for distance $x$ provided that the dimer is (permanently) in the state $s$.

Substituting these expressions into eqn~(\ref{p0}) and (\ref{p1}) and integrating over the variable $x$, one finds that the probabilities $\pi_s$ obey classical master equations for a two-level system, 
\begin{align}
\frac{d\pi_0}{dt}=w_1\pi_1-w_0\pi_0,
\end{align}
and
\begin{align}
\frac{d\pi_1}{dt}=w_0\pi_0-w_1\pi_1.
\end{align}

Here $w_0$ and $w_1$ are effective rates of transitions between the states given by 
\begin{align}
w_0=\int_{-\infty}^{\infty}dx\, u_0(x)p(x|s=0),
\end{align}
and
\begin{align}
w_1=\int_{-\infty}^{\infty}dx\,u_1(x)p(x|s=1).
\end{align}

The involved probability distributions in the statistically stationary  case are
\begin{align}
p(x|s=0)=\sqrt {\frac{k_0}{2\pi k_{\rm B}T }}\exp \left[-\frac{k_0}{2k_{\rm B}T}(x -\ell_0)^2\right] ,
\end{align}
and
\begin{align}
p(x|s=1)=\sqrt {\frac{k_1}{2\pi k_{\rm B}T }}\exp \left[-\frac{k_1}{2k_{\rm B}T}(x -\ell_1)^2\right].
\end{align}

If the transition windows are narrow, approximations in eqn~(\ref{delta}) can furthermore be used, so that we obtain 
\begin{align}
w_0=\nu_0 p(x=\ell_0|s=0),~~~~~w_1=\nu_1 p(x=\ell_1|s=1).
\end{align}

Thus, using the above expressions for distance distributions, we finally get
\begin{align}
w_0=2\rho v_0 \sqrt {\frac{k_0}{2\pi k_{\rm B}T }},
\end{align}
and
\begin{align}
w_1=2\rho v_1 \sqrt {\frac{k_1}{2\pi k_{\rm B}T }}.
\end{align}

In the steady state, the probabilities are 
\begin{align}
\pi_0=\frac{w_1}{w_0+w_1},~~~~~\pi_1=\frac{w_0}{w_0+w_1}.
\end{align}

\section{Average force dipole}
\label{app:m}

Let us consider the second statistical moment $\langle x^2 \rangle$. In a steady state, its time derivative is zero. On the other hand, by using eqn~(\ref{p0})--(\ref{p1}) and integrating by parts, we find 
\begin{align}
& \frac{d\langle x^2 \rangle}{dt} 
\nonumber \\
& = 2 \gamma \int_{-\infty}^\infty dx\,\Big [k_0x(\ell_0-x)p_0(x)+k_1x(\ell_1-x)p_1(x)\Big]
\nonumber \\
& +2\gamma k_{\rm B}T \int_{-\infty}^\infty  dx\, \Big [p_0(x) +p_1(x) \Big] 
\nonumber \\
& =2\gamma \langle m \rangle + 2\gamma k_{\rm B}T =0.
\end{align} 
Thus, we straightforwardly obtain that, for an active dimer in any statistically steady state, $\langle m \rangle = - k_{\rm B}T$.

Note that here and also in the equations below, the integration limits over $x$ are taken as $+\infty$ and $-\infty$. The actual limits are automatically selected by probability distributions $p_0(x)$ and $p_1(x)$.

\section{Force-dipole correlation function}
\label{app:FP}

Introducing 
\begin{align}
\mathbf{p}(x,t)= 
\left(
\begin{tabular}{c}
$p_0(x,t)$  \\ $p_1(x,t)$
\end{tabular}
\right),
\end{align}
we can write the system of two master equations (\ref{p0}) and (\ref{p1}) concisely as 
\begin{align}
\frac{d\mathbf{p}}{dt}=-\mathbf{\hat{L}}\mathbf{p},
\label{master}
\end{align}
where 
\begin{align}
\mathbf{\hat{L}}=\left(
\begin{tabular}{cc}
$\hat{L}_{00}$& $\hat{L}_{01}$ \\ $\hat{L}_{10}$&$\hat{L}_{11}$
\end{tabular}
\right),
\end{align}
and 
\begin{align}
\hat{L}_{00}=\gamma k_0 \frac{\partial}{\partial x} (\ell_0-x)-\gamma k_{\rm B} T \frac{\partial^2}{\partial x^2}+u_0(x),
\end{align}
and
\begin{align}
\hat{L}_{11}=\gamma k_1 \frac{\partial}{\partial x} (\ell_1-x)-\gamma k_{\rm B} T \frac{\partial^2}{\partial x^2}+u_1(x),
\end{align}
and
\begin{align}
\hat{L}_{01}=-u_1(x),~~~~~\hat{L}_{10}=-u_0(x).
\end{align}

The general solution of eqn~(\ref{master}) is 
\begin{align}
p_s(x,t)=\sum_{n=0}^\infty A_n q_s^{(n)} (x)e^{-\lambda_n t}+{\rm c.c.}
\end{align}
where $\lambda_n$ and $\mathbf{q}^{(n)}$ are eigenvalues and eigenvectors of the linear operator $\mathbf{\hat{L}}$,
\begin{align}
\mathbf{\hat{L}}\mathbf{q}^{(n)}=\lambda_n \mathbf{q}^{(n)},
\end{align}
and decomposition coefficients $A_n$  are determined by initial conditions.

Because the master equation must have a stable stationary solution, the operator $\mathbf{\hat{L}}$ should always possess a zero eigenvalue $\lambda_0=0$ and, furthermore, condition $\text{Re}\,\lambda_n>0$ should hold for all other eigenvalues $n$.\cite{risken}
Generally, the eigenvectors can be ordered according to the increase of $\text{Re}\,\lambda_n$ (and therefore we can enumerate the eigenvalues in such a way that  $0<\text{Re}\,\lambda_1\leq \text{Re}\,\lambda_2\leq \text{Re}\,\lambda_3\leq...$).
The stationary probability distribution $\mathbf{\bar{p}}(x)$ coincides with the eigenvector $\mathbf{q}^{(0)}(x)$.

The conditional probability $G(x,s,t|x_0,s_0)$ gives the probability to find the dimer in various states $(x,s)$ at time $t$ provided that it was in the state $(x_0,s_0)$ at time $t=0$. It represents a special solution of the master equation (\ref{master}) given by
\begin{align}
G(x,s,t|x_0,s_0)=\sum_{n=0}^\infty a_n(x_0,s_0) q_s^{(n)} (x)e^{-\lambda_n t}+{\rm c.c.}
\label{conditional}
\end{align}
where $a_n(x_0,s_0)$ are the coefficients of decomposition of this initial condition over eigenvectors $\mathbf{q}^{(n)}$.

The force dipole $m$ depends on the distance $x$ between the domains and on the dimer state $s$, i.e., $m(t)=m(x(t),s(t))$. Therefore, in the statistically stationary state we have 
\begin{align}\label{mm}
 \langle m(t)m(0) \rangle &= \sum_{s,s_0=0,1}\int_{-\infty}^\infty dx_0\int_{-\infty}^\infty dx\, m(x_0,s_0)m(x,s) \nonumber \\
 &\times \bar{p}_{s_0}(x_0)G(x,s,t|x_0,s_0).
\end{align} 
By using eqn~(\ref{conditional}) and (\ref{mm}), we find that, in the statistically stationary state, the correlation function of force dipoles is \begin{align}
C(t)=\langle m(t)m(0) \rangle-\langle m^2 \rangle= \sum_{n=1}^\infty B_n e^{-\lambda_n|t|} + {\rm c.c.}
\end{align}
where the complex coefficients $B_n$ are 
\begin{align}
B_n & =\sum_{s,s_0=0,1}\int_{-\infty}^\infty dx_0\int_{-\infty}^\infty dx\, m(x_0,s_0)m(x,s)\nonumber \\
&\times \bar{p}_{s_0}(x_0)a_n(x_0,s_0) q_s^{(n)} (x).
\end{align}

If we retain in this decomposition only the first, most slowly decaying term, this yields 
\begin{align}
C(t)\approx B_1e^{-\lambda_1|t|}+{\rm c.c.}=\frac{C(0)}{\cos\alpha}e^{-\Gamma |t|}\cos(\Omega |t|-\alpha).
\end{align}
Therefore, the normalized correlation function is
\begin{align}
\frac{C(t)}{C(0)}=\frac{1}{\cos\alpha}e^{-\Gamma|t|}\cos(\Omega|t|-\alpha),
\label{normalized}
\end{align}
where $\Gamma=\text{Re}\,\lambda_1$, $\Omega=\text{Im}\,\lambda_1$, and $B_1=C(0)e^{i\alpha}/\cos\alpha$.

Our numerical simulations, described in Section~\ref{sec:numerical}, have shown that, in the regimes approaching a deterministic oscillatory dimer, the correlation functions of force dipoles could be well fitted to the above dependence. This suggests that contributions from the higher, more rapidly decaying relaxation modes $n>1$ have been indeed relatively small. As generally known,\cite{kuramoto} noisy oscillators possess a slowly relaxing mode that corresponds to diffusion of the oscillation phase. It can be expected that, under chosen conditions, such a mode has been dominating the correlation functions for oscillatory dimers.

\section{Dependence on orientational correlation time}
\label{app:DE}

Suppose that the force-dipole correlation function $C(t)$ and the orientational correlation function $\sigma(t)$ are given by eqn~(\ref{normalized}) and (\ref{rotation}). 
By taking the integral in eqn~(\ref{chi}), we find 
\begin{align}
\chi=\frac{1/\tau_{\rm rot} +\Gamma+\Omega\tan\alpha}{(1/\tau_{\rm rot}+\Gamma)^2+\Omega^2}\langle \Delta m^2 \rangle
\label{chi1}.
\end{align}
This yields a non-monotonous dependence of $\chi$ on the orientational correlation time. 
If the phase shift $\alpha$ is small and can be neglected (cf.\ Fig.~\ref{fig7}), the maximum value $\chi_{\rm max}$ is reached at $\tau_{\rm rot}=(\Omega-\Gamma)^{-1}$ and we have
\begin{align}
\frac{\chi_{\rm max}}{\chi_\infty}=\frac{\Gamma^2+\Omega^2}{2\Gamma\Omega},
\end{align}
where
\begin{align}
\chi_\infty=\frac{\Gamma}{\Gamma^2+\Omega^2}\langle \Delta m^2 \rangle,
\end{align}
is the limit of $\chi$ when $\tau_{\rm rot}\gg\Gamma^{-1}$ and $\tau_{\rm rot}\gg\Omega^{-1}$.

These results allow us to discuss how the diffusion enhancement would depend on the orientational correlational time $\tau_{\rm rot}$, not assuming that it is much shorter than the correlation time for force dipoles.
If the approximation in eqn~(\ref{damped}) holds, diffusion enhancement is determined by eqn~(\ref{enhancement}) where $\chi$ is given by eqn~(\ref{chi1}).
The diffusion enhancement depends non-monotonously on the orientational correlation time. It increases linearly with $\tau_{\rm rot}$ at short times, then reaches a maximum at $\tau_{\rm rot}=(\Omega-\Gamma)^{-1}$ and finally saturates at large orientational correlation times.
For example, if we take the values $\Gamma\approx1/(2\tau_0)$ and $\Omega\approx\pi/(3\tau_0)$ corresponding to substrate saturation in Fig.~\ref{fig7}, the maximum diffusion enhancement would be reached at $\tau_{\rm rot}=1.8\tau_0$ and, at the maximum, it will be larger by about 30 percent than in the limit $\tau_{\rm rot}\gg\tau_0$.

\bibliography{ref} 
\bibliographystyle{rsc} 

\end{document}